\documentclass[aps,fleqn]{revtex4}

\newcommand{\bed}{\[}
\newcommand{\eed}{\]}
\newcommand{\beq}{\begin{equation}}
\newcommand{\eeq}{\end{equation}}
\newcommand{\beqa}{\begin{eqnarray}}
\newcommand{\eeqa}{\end{eqnarray}}
\newcommand{\ket} [1] {\vert #1 \rangle}
\newcommand{\bra} [1] {\langle #1 \vert}

\newcommand{\gras}[1]{\bold{#1}}

\newcommand{\Tr}{\mathop{\mathrm{Tr}}}
\newcommand{\tr}{\mathop{\mathrm{tr}}}

\usepackage{graphicx}
\usepackage[mathscr]{eucal}
\usepackage{amsmath}
\usepackage{amssymb}
\usepackage{epstopdf}
\usepackage{epsfig}
\begin{document}

\title{Generalised Asymmetric Quantum Cloning Machines}
\author{S. Iblisdir$^1$, A. Ac\'in$^2$ and N. Gisin$^1$}
\address{$^1$GAP-Optique, University of Geneva,
20 rue de l'Ecole-de-M\'edecine, CH-1211, Switzerland \\
$^2$ Institut de Ci\`encies Fot\`oniques, Jordi Girona 29,
08034 Barcelona, Spain}

\date{\today}

\begin{abstract}
We study machines that take $N$ identical replicas of a pure qudit state as input and output  a set of $M_A$ clones of a given fidelity and another set of $M_B$ clones of another fidelity. The trade-off between these two fidelities is investigated, and numerous examples of optimal $N \to M_A+M_B$ cloning machines are exhibited using a generic method. A generalisation to more than two sets of clones is also discussed. Finally, an optical implementation of some such machines is proposed. This paper is an extended version of [xxx.arxiv.org/abs/quant-ph/0411179].
\end{abstract}

\maketitle

\section{Introduction}\label{section:intro}

The no-cloning theorem \cite{woot82} states that it is in general impossible to perfectly clone the state of a quantum system. This no-go result is a typical feature of quantum information, and is deeply related to quantum error correction \cite{preskill}, or the impossibility of super-luminal signalling \cite{diek82}. But the no-cloning theorem can also be turned to a valuable resource as demonstrated by quantum cryptography \cite{gisi02}.

Although perfect cloning is forbidden by quantum mechanics, it is of fundamental importance to analyse how well we can approximately copy a quantum state into several quantum systems in order to understand how quantum information distributes. The simplest instance of this problem is the duplication of the pure state of a qubit, and has been considered in \cite{buze96} (a qubit is a two-level quantum system).  Many generalisations and variants have followed. Among them are the issue of producing $M$ clones of a qubit state from $N$ input replicas \cite{gisi97} and its generalisation to qudits ($d$-level quantum systems) \cite{wern98,keyl98}, cloning from non identical input states \cite{fiur02}, or non-universal cloning \cite{dari01}. Cloning of continuous variable systems has also been considered \cite{cerf00}.

This paper deals with universal $N \to M_A+M_B+ \ldots$ asymmetric cloning of pure qudit states. That is, we are given $N$ identical replicas of an unknown pure qudit state, and we want to produce $M_A$ approximate clones all with a same fidelity, say $F^A$, $M_B$ clones, all with a same fidelity $F^B$, $\dots$. We are interested in the trade-off between the qualities of the various sets of clones. Such machines have already been examined in \cite{cerf00:asym}, but only $1 \to 1+1$ cloning was considered, and optimality was only proven in the case of qubits. $N \to M_A+M_B$ cloning machines have at least two interesting applications. First, in the case where $M_A=N$ and $M_B \to \infty$, we get a machine that allows to study the trade-off between the acquisition of knowledge about the state of a quantum system and the disturbance undergone by this system. Second, some $N \to M_A+M_B$ cloning machines have been proven to be a useful tool when investigating the security of some quantum key distribution schemes \cite{nied05,acin04,curt03}.

Here is a summary of our results.

\begin{itemize}

\item A generic method to get optimal $N \to M_A+M_B$ cloning machines of pure qudit states is prescribed for any values of $N, M_A,M_B,d$, and a  'natural' conjecture of what cloning machines this method should always produce is proposed . This conjecture has been supported by all the cases we have examined.  These results are essentially the content of  Section \ref{section:algebra}.

\item With the aforementioned method, optimal asymmetric  cloning of qubits has been investigated in the following cases: $1 \to 1+n$ cloning of
qubits  (see section \ref{section:qubitetmesure}) and $n \to n+1$ cloning of qubits (see Section \ref{subsection:twoplusone}).

\item Optimal $1 \to 1+1+1$ cloning of qubits is analysed (Section\ref{sec:unplusunplusun}).

\item A probabilistic implementation of some such asymmetric machines is proposed. (Section \ref{sec:implementations}).

\end{itemize}

%%%%%%%%%%%%%%%%%%%%%%%%%%%%%%%%
\section{Optimal asymmetric cloning maps}\label{section:algebra}
%%%%%%%%%%%%%%%%%%%%%%%%%%%%%%%%

Before going into the details of cloning maps, let us sketch the idea behind the subsequent analysis. Let us consider the simplest case of $1 \to 2$ universal cloning of qubits. Any one-qubit state can be represented by a unit vector $\gras{n}$ on the surface of a sphere, the so-called Bloch vector on the Bloch sphere. The effect of an optimal symmetric cloning machine is to produce two output clones with reduced Bloch vectors $\eta \gras{n}$ ($\eta < 1$). That is, the cloning machine merely shrinks the input Bloch vector, but doesn't affect its orientation.  This "isotropy" property results from the fact that no state is preferred by an optimal universal cloning machine. Therefore the quality of the cloning process is completely characterised by the shrinking factor $\eta$. As demonstrated in \cite{keyl98}, a similar argument applies for $N \to M$ cloning of qudits: the quality of the clones is fully characterised by  some quantity $\omega$, and the whole problem of optimal cloning amounts to extremise this quantity. As we will observe below, this argument still applies for \emph{asymmetric} cloning. Consider $N \to M$ cloning and let $M_A+M_B=M$ denote a partition of $M$. If we require $M_A$ clones to be of the same quality, and $M_B$ clones to be of the same quality, the quality of the first (respectively second) set of clones is fully characterised by some quantity $\omega^A$ (respectively $\omega^B$). Therefore, from an algebraic point of view, the main problem posed by optimal asymmetric cloning is to find some tight  relation between $\omega^A$ and $\omega^B$. 

The object of this section is threefold: (i) to cast the cloning problem into purely algebraic terms, (ii) to give a quantitative relation between the qualities of the two sets of clones, (iii)  to show how to get optimal cloning machines.

%%%%%%%%%%%%%%%%%%%%%%%
\subsection{Cloning maps and figures of merit}
%%%%%%%%%%%%%%%%%%%%%%%

An asymmetric  cloning device would be an operation taking $N$ identical replicas of an unknown state  and outputing $M=M_A+M_B$ approximate clones. In Schr\"odinger picture, this operation is described by a trace-preserving completely positive map \cite{keyl02} 
\bed
T':\mathscr{B}(\mathscr{H}^{\otimes N}) \to \mathscr{B}(\mathscr{H}^{\otimes M}).
\eed
$\mathscr{H}=\gras{C}^d$ denotes the Hilbert space of a single qudit, $\mathscr{H}^{\otimes N}$ its $N$-fold tensor product, and $\mathscr{B}(\mathscr{H}^{\otimes N})$ denotes the space of bounded operators over $\mathscr{H}^{\otimes N}$. $T'$ makes states evolve and leaves operators invariant . Equivalently, cloning can be described in Heisenberg picture by a unital ($T(\textbf{1})=\textbf{1}$) completely positive map $T:\mathscr{B}(\mathscr{H}^{\otimes M}) \to \mathscr{B}(\mathscr{H}^{\otimes N})$. $T$ makes operators evolve and states are now left invariant. The two pictures are equivalent and are related by the identity: 
\beq 
\tr(\rho T(\mathscr{O}))=\tr(\mathscr{O} T'(\rho)), 
\eeq
for all operator $\mathscr{O} \in \mathscr{B}(\mathscr{H}^{\otimes M})$ and for all density operator $\rho \in \mathscr{B}(\mathscr{H}^{\otimes N})$. Note that since all input systems are of the form $P_{\psi}^{\otimes N} \equiv \ket{\psi}\bra{\psi}^{\otimes N}$, $\ket{\psi} \in \mathscr{H}$, we can without loss of generality consider the input Hilbert space to be only the symmetric subspace of $\mathscr{H}^{\otimes N}$: $\mathscr{H}^{\otimes N}_{+}$. So, cloning can be described by a unital cp-map 
\beq
T:\mathscr{B(H}^{\otimes M}) \to \mathscr{B(H}^{\otimes N}_{+}).
\eeq

There are essentially two options to quantify the quality of the clones. The first one consists in considering global figures of merit. That is if $\psi$ denotes the  input state to clone, one
can consider the quantity 
\beqa\label{eq:defglobalfid}
F^A_{\textrm{all}}(T) & = &  \textrm{min}_{\psi}
\bra{\psi^{\otimes M_A}} \Tr {}_{\mathscr{H}^{\otimes M_B}}
T'(P_{\psi}^{\otimes N}) \ket{\psi^{\otimes M_A}} \nonumber \\
& = &  \textrm{min}_{\psi}  \bra{\psi^{\otimes N}}
(T(P_{\psi}^{\otimes M_A} \otimes \gras{1}^{\otimes M_B})
\ket{\psi^{\otimes N}}, 
\eeqa 
to quantify the first set of $M_A$ clones, where $\Tr {}_{\mathscr{H}^{\otimes M_B}}$ denotes the partial trace over the second set of clones. A similar expression holds for the global fidelity of  the second set of clones, $F^B_{\textrm{all}}(T)$.

Alternatively, one can consider  single-copy fidelities:
\beqa\label{eq:deflocalfid}
F^A(T) & = &  \textrm{min}_{\psi \in \mathscr{H}} \textrm{min}_{1 \leq k \leq M_A}
\bra{\psi} \Tr {}_{k} T'(\ket{\psi^{\otimes N}}\bra{\psi^{\otimes N}})
\ket{\psi} \nonumber \\
& = &  \textrm{min}_{\psi \in \mathscr{H},k}  \bra{\psi^{\otimes
N}} (T(\gras{1}^{\otimes k-1} \otimes \ket{\psi} \bra{\psi}
\otimes \gras{1}^{\otimes M_A+M_B-k})\ket{\psi^{\otimes N}}, 
\eeqa
for the first set of $M_A$ clones. In this expression, $\Tr{}_ {k}$ denotes the partial trace over all clones but the $k$-th in the first set. A similar expression holds for the single-clone
fidelity of  the second set of clones, $F^B(T)$.

Even in the symmetric case, it is highly non-trivial to prove that a quantum cloning machine which is optimal for one figure of merit is optimal for the other \cite{keyl98}. In this work we choose to consider single-clone fidelities because it is more relevant than the global fidelities when one considers connections of cloning with
other tasks, such as state estimation (see Section \ref{section:qubitetmesure}), or quantum cryptography \cite{acin04}.

Considering the first set of clones, it is obvious that $F^A(T)$ lies between the fidelity obtained when preparing the clones in a random state, and the fidelity of the optimal $N
\to M_A$ symmetric cloning machine $F^A_{\textrm{sym}}(N,M_A)$. The problem of finding optimal $N \to M_A+M_B$ asymmetric cloning can now be clearly formulated: To maximise $F^B(T)$ for a given value of $F^A(T)$ between $1/d$ (the fidelity of a machine producing outputs in a random state) and $F^A_{\textrm{sym}}(N,M_A)$ (the fidelity of an optimal $N \to M_A$ symmetric machine).

Before proceeding any further, let us fix some notations. These notations are similar to those in the paper of Keyl and Werner \cite{keyl98}, so as to make the connection between their work and ours as transparent as possible. $\textrm{U}(d)$ will denote the  (compact Lie) group of unitary $d \times d$ matrices, and $\textrm{SU}(d)$ will denote the subgroup of $\textrm{U}(d)$ of elements whose determinant equals $1$. The representations of $\textrm{U}(d)$ will be denoted $\pi_{\alpha}$.
Of particular importance are: (i) the natural representation, i.e. the representation given by the elements of the group themselves, which we will simply denote as $\pi$, (ii) the $N$-fold tensor product of $\pi$, $\pi^{\otimes N}$ and (iii) the irreducible representation given by the restriction of $\pi^{\otimes N}$ to the symmetric subspace $\mathscr{H}^{\otimes N}_+$ of
$\mathscr{H}^{\otimes N}$: $\pi_N^+$. Thus, if $\psi_0 \in \gras{C}^{d}$ denotes some fiducial state, the set of input states to clone reads 
\bed
\{\pi_N^+(u) \ket{\psi_0}^{\otimes N}: \; u \in\textrm{U}(d) \}.
\eed
In the special case $d=2$, as is well known, the irreducible representations are labelled by half-integer positive numbers. Accordingly, these irreducible representations will be denoted $\pi_0,\pi_{1/2}$, etc.
$\textrm{u}(d)$ will denote the Lie algebra associated to $\textrm{U}(d)$, i.e. $\textrm{u}(d)$ is the Lie algebra of antihermitian matrices, and $\textrm{su}(d)$, the
algebra consisting of traceless elements of $\textrm{u}(d)$. The irreducible representations of $\textrm{u}(d)$ will be denoted $\partial\pi_{\alpha}$. Let $X$ denote an element of
$\textrm{u}(d)$ and let $\{ e^{t X} \in \textrm{U}(d): t \in \gras{R} \}$ denote the associated one-parameter subgroup of $\textrm{U}(d)$. We have 
\bed 
\pi_{\alpha}(e^{t X})=e^{t \partial
\pi_{\alpha}(X)}. 
\eed 
Also note that the representation of $\partial \pi^{\otimes M}$ can be expressed very simply from the natural representation $\partial \pi$: 
\bed
\partial \pi^{\otimes M}(X)=\sum_{i=1}^M \gras{1}^{\otimes i-1} \otimes \partial \pi(X)
\otimes \gras{1}^{\otimes M-i}.
\eed

%%%%%%%%%%%%%%%%%%%%%%%%%%%%%
\subsection{A natural conjecture}\label{section:conjecture}
%%%%%%%%%%%%%%%%%%%%%%%%%%%%%

The optimal symmetric cloning map reads (in Schr\"odinger picture) \cite{wern98}: \beq\label{eq:clonsym} T'_{\textrm{sym}}: \mathscr{B}(\mathscr{H}^{\otimes N}_+) \to \mathscr{B}(\mathscr{H}^{\otimes M}): \rho^{\otimes N} \to \frac{d[N]}{d[M]} S_M (\rho^{\otimes N} \otimes \gras{1}^{\otimes M-N}) S_M, \eeq where $d[N]=\textrm{dim}\mathscr{H}^{\otimes N}_+$ (the constant $d[N]/d[M]$ ensures that the map is trace-preserving). $S_M$ is the projector onto the symmetric subspace $\mathscr{H}^{\otimes M}_+$. The interpretation of this map is quite intuitive: $M-N$ states containing no (quantum) information are appended to the input, and the resulting state is symmetrised.

Note that  the representation $\pi^{\otimes M}$ decomposes as $\pi^{\otimes M}=
\pi_M^+ \oplus \pi_{\textrm{rest}}$, where $\pi_{\textrm{rest}}$ is some representation containing no representation equivalent to $\pi_M^+$ and acting on some space $\mathscr{H}_{\textrm{rest}}$ \cite{zhelobenko}. Accordingly, we have $\gras{1}^{\otimes M}=S_M+ \gras{1}_{\textrm{rest}}$, where $\gras{1}_{\textrm{rest}}$ denotes the identity over $\mathscr{H}_{\textrm{rest}}$. Thus, what the cp-map (\ref{eq:clonsym}) suggests is that optimality is achieved by keeping only the component of the decomposition corresponding to the symmetric subspace.

In the case of asymmetric cloning, we have to consider the
decomposition  $\pi^{\otimes M_A} \otimes \pi^{\otimes M_B}
\approx (\pi^+_{M_A} \otimes \pi^+_{M_B}) \oplus (\pi^+_{M_A}
\otimes \pi_{\textrm{rest'}}) \oplus (\pi_{\textrm{rest}} \otimes
\pi^+_{M_B}) \oplus (\pi_{\textrm{rest}} \otimes
\pi_{\textrm{rest'}})$. According to this decomposition, we have
$\gras{1}^{\otimes M}=\gras{1}^{\otimes M_A} \otimes \gras{1}^{
\otimes M_B}=S_{M_A} \otimes S_{M_B}+ S_{M_A} \otimes
\gras{1}_{\textrm{rest'}}+\gras{1}_{\textrm{rest}} \otimes
S_{M_B}+ \gras{1}_{\textrm{rest}} \otimes
\gras{1}_{\textrm{rest'}}$.

Our conjecture is that only the piece $\pi^+_{M_A} \otimes
\pi^+_{M_B}$ should be considered in this decomposition. More
precisely, let $\pi^+_{M_A} \otimes \pi^+_{M_B} \approx
\oplus_{\gamma} \pi_{\gamma}$ denote the decomposition of
$\pi^+_{M_A} \otimes \pi^+_{M_B}$ into irreducible components, and
let $E_{\gamma}$ denote the projector associated to the
irreducible component $\pi_{\gamma}$. We conjecture that optimal asymmetric cloning
machines should be of the form: \beq\label{eq:conjecture}
T'_{\textrm{asym}}: \mathscr{B}(\mathscr{H}^{\otimes N}_+) \to
\mathscr{B}(\mathscr{H}^{\otimes M}): \rho^{\otimes N} \to V^*
(\rho^{\otimes N} \otimes \gras{1}^{\otimes M-N}) V, \eeq where
$V$ is a linear combination of projectors $E_{\gamma}$. This
conjecture is supported by \emph{all} asymmetric machines we have considered.

%%%%%%%%%%%%%%%%%%%%%%%%%%%%%%%%%%%%%%%%
\subsection{How to get optimal asymmetric cloning maps}\label{section:generic}
%%%%%%%%%%%%%%%%%%%%%%%%%%%%%%%%%%%%%%%%

We now describe a general recipe to get optimal asymmetric cloning machines. This section summarises the results obtained in Appendix \ref{section:symoptimalmap}. $N \to M_A+M_B$ cloning of qudits is achieved by a cp-map $T: \mathscr{B}(\mathscr{H}^{\otimes M}) \to \mathscr{B}(\mathscr{H}_N^+)$, which decomposes as 
\beq 
T= \sum_{\alpha_1 \in D(\pi^{ \otimes M_A})} \;
\sum_{\alpha_2 \in D(\pi^{\otimes M_B})} \;
\sum_{\beta \in D_N(\pi_{\alpha_1} \otimes \pi_{\alpha_2})}
\; r(\alpha_1,\alpha_2,\beta) \;T(\alpha_1,\alpha_2,\beta), \\
\eeq 
where we define $D(\pi_Z)=\{a: \pi_a \subset \pi_Z\}$ and $D_N(\pi_Z)=\{ a: \pi_N^+ \subset \pi_Z \otimes \pi_a \}$. The quantities $r(\alpha_1,\alpha_2,\beta)$ satisfy
\beq\label{eq:constraintonr} 
r(\alpha_1,\alpha_2,\beta) \geq 0,
\hspace{1cm} \sum_{\alpha_1,\alpha_2,\beta}
r(\alpha_1,\alpha_2,\beta)=1. 
\eeq

The (single-clone) fidelities are essentially fixed by two quantities, $\omega^A(T)$ and $\omega^B(T)$, analogous to the shinrking factor $\eta$ discussed at the beginning of Sect.\ref{section:algebra}. We have
\beqa
F^A(T) &=& \frac{1}{d}(1+\frac{N}{M_A} \omega^A(T) (d-1)), \\
F^B(T) &=& \frac{1}{d}(1+\frac{N}{M_B} \omega^B(T) (d-1)). 
\eeqa

The quantities $\omega^A(T)$ and $\omega^B(T)$ are decomposed according to the convex decomposition of $T$ into irreducible summands: 
\beqa
\omega^A(T) &=& \sum_{\alpha_1 \in D(\pi^{ \otimes M_A})}
\; \sum_{\alpha_2 \in D(\pi^{\otimes M_B})}
\;\sum_{\beta \in D_N(\pi_{\alpha_1} \otimes \pi_{\alpha_2})}
\; r(\alpha_1,\alpha_2,\beta) \;\omega^A(T(\alpha_1,\alpha_2,\beta)), \\
\omega^B(T) &=& \sum_{\alpha_1 \in D(\pi^{\otimes M_A})} \;
\sum_{\alpha_2 \in D(\pi^{\otimes M_B})} \;
\sum_{\beta \in D_N(\pi_{\alpha_1} \otimes \pi_{\alpha_2})} \;
r(\alpha_1,\alpha_2,\beta) \;\omega^B(T(\alpha_1,\alpha_2,\beta)).
\eeqa

The quantities $\omega^A(T(\alpha_1,\alpha_2,\beta)$ are given by
\beq\label{eq:omegacasimir} 
\omega^A_{\alpha_1,\alpha_2,\beta}= \frac{1}{2} \sum_{a \in
D(\pi_{\alpha_2} \otimes \pi_{\beta})} |\lambda_a|^2 
\{ 1+\frac{C_2(\pi_{\alpha_1})- C_2(\pi_a)}{ C_2(\pi_N^+)} \}, 
\eeq
where $C_2(\pi_{a})$ denotes the Casimir number associated to the irreducible representation $\pi_a$. The quantities $\lambda_a$ satisfy 
\beq\label{eq:sumlambdasquared} 
\sum_a |\lambda_a|^2=1.
\eeq 
Similarly, 
\beq
\omega^B_{\alpha_1,\alpha_2,\beta}=\frac{1}{2} \sum_{b \in D(\pi_{\alpha_1} \otimes \pi_{\beta})}
|\mu_b|^2 \{ 1+ \frac{C_2(\pi_{\alpha_2})- C_2(\pi_b)}{
C_2(\pi_N^+)} \},
\eeq
 with $\sum_b |\mu_b|^2=1$.

 The quantities $\lambda_a$ and $\mu_b$ are related by 
 \beq\label{eq:relationmulambda}
\mu_b=\frac{1}{\textrm{dim}\mathscr{H}^{\otimes N}_+}\sum_a
\lambda_a  \tr W^*_{b} V_a. 
\eeq 
$V_a$ and $W_b$ are intertwining operators defined as follows. Let $X$ denote an auxiliary space supporting the representation $\beta$. 
Consider the decomposition theory of $\pi_{\alpha_1}^A \otimes \pi_{\alpha_2}^B \otimes \pi_{\beta}^X \approx \pi_N^+ \oplus \pi_{\textrm{rest}}$, where $\pi_{\textrm{rest}}$ contains no copy of $\pi_N^+$. $V_a$ and $W_b$ are defined as the unique isometries such that
\beqa\label{eq:intertwiningprop}
(\pi_{\alpha_1}^A \otimes \pi_a^{BX}) V_a &=& V_a \pi_N^+, \\
(\pi_{\alpha_2}^B \otimes \pi_b^{AX}) W_b &=& W_b \pi_N^+. 
\eeqa
It is now clear that the problem of finding optimal cloning machines is a constrained optimisation problem: We have to maximize $\omega^B(T)$ for a fixed value of $F^A(T)$ taking into account the constraint (\ref{eq:constraintonr})-(\ref{eq:sumlambdasquared})-(\ref{eq:relationmulambda}).

In the next section, we apply the recipe just described to treat some concrete examples.

 %%%%%%%%%%%%%%%%%%%%%%%
\section{Some asymmetric cloning machines}
%%%%%%%%%%%%%%%%%%%%%%%

%%%%%%%%%%%%%%%%%%%%%%%%%%%%%%%%%%%%%%%%%%%
\subsection{The simplest example: $1 \to 1+1$ cloning of qubits}\label{section:simplest}
%%%%%%%%%%%%%%%%%%%%%%%%%%%%%%%%%%%%%%%%%%%

Although $1 \to 1+1$ cloning machines of qubits have been extensively studied \cite{cerf00:asym}, it is instructive to revisit them in order to illustrate as simply as possible the foregoing analysis.

Adopting the standard convention  of denoting irreducible representations of SU$(2)$ by half-integer numbers, we have here $D(\pi^{\otimes M_A})=D(\pi^{\otimes M_B})=\{1/2\}$, and $D_1(1/2\otimes 1/2)=\{ \beta : 1/2 \subset 1/2 \otimes 1/2 \otimes \beta\}=\{3/2,1/2\} $, where use was made of the Clebsch-Gordan series 
\bed 
j_1 \otimes j_2 \approx |j_1-j_2| \oplus \ldots \oplus (j_1+j_2), 
\eed
for the decomposition of the tensor product of two irreducible
representations. Accordingly, 
\beqa 
\omega^A(T) &=& r(1/2,1/2,3/2)
\omega^A(T(1/2,1/2,3/2))+r(1/2,1/2,1/2)
\omega^A(T(1/2,1/2,1/2)), \\
\omega^B(T) &=& r(1/2,1/2,3/2) \omega^B(T(1/2,1/2,3/2))+r(1/2,1/2,1/2)
\omega^B(T(1/2,1/2,1/2)).
\eeqa

With $C_2(j)=j(j+1)$, we have:

\bed 
\omega^A(T(1/2,1/2,3/2))=\omega^B(T(1/2,1/2,3/2))=
\frac{1}{2}(1+\frac{C_2(1/2)-C_2(1)}{C_2(1/2)}) =-1/3. 
\eed 
So the map $T(1/2,1/2,3/2)$ is useless for cloning, since the quantities $\omega^A$ and $\omega^B$ it yields are worse than for a map which would consit of preparing the clones in a random state ($\omega^A_{\textrm{random}}=\omega^B_{\textrm{random}}=0$). Let us consider the other map, $T(1/2,1/2,1/2)$. We have 
\beqa
\omega^A(T(1/2,1/2,1/2)) &=&
\frac{1}{2}(|\lambda_0|^2 (1+\frac{C_2(1/2)-C_2(1/2)}{C_2(0)})+
|\lambda_1|^2 (1+\frac{C_2(1/2)-C_2(1)}{C_2(1/2)}) \nonumber \\
&=& \frac{1}{2}(2|\lambda_0|^2-\frac{2}{3}|\lambda_1|^2). 
\eeqa
Similarly, 
\beq
\omega^B(T(1/2,1/2,1/2))=\frac{1}{2}(2|\mu_0|^2-\frac{2}{3}|\mu_1|^2).
\eeq 
Let us now work out the relation between the coefficients $\lambda_0, \lambda_1$ and $\mu_0,\mu_1$. In turn, this relation will give us the trade-off between the fidelities of the clone A
and the clone B. This relation involves four isometric interwiners: $V_0, V_1, W_0$ and $W_1$. Explicitly, we have 
\beq\label{eq:simplestintertwiner1} 
V_0 = \ket{({1/2}^A 0^{BX}) 1/2, m }\bra{1/2, m}=
 C^{(1/2 s)(0 t )}_{(1/2 m)} C^{(1/2 u)(1/2 v )}_{(0 t)}\ket{1/2, s}_A \ket{1/2, u}_B \ket{1/2, v}_X \bra{1/2, m}, \eeq
\beq\label{eq:simplestintertwiner2} 
 V_1 = \ket{({1/2}^A  1^{BX}) 1/2, m }\bra{1/2, m}=
C^{(1/2 s)(1 t )}_{(1/2 m)} C^{(1/2 u)(1/2 v )}_{(1 t)}\ket{1/2, s}_A \ket{1/2, u}_B \ket{1/2, v}_X \bra{1/2, m}, 
\eeq
\beq\label{eq:simplestintertwiner3} 
W_0 = \ket{({1/2}^B  0^{AX}) 1/2, m }\bra{1/2, m}=
C^{(1/2 s)(0 t )}_{(1/2 m)} C^{(1/2 u)(1/2 v )}_{(0 t)}\ket{1/2, u}_A \ket{1/2, s}_B \ket{1/2, v}_X \bra{1/2, m},
\eeq
\beq\label{eq:simplestintertwiner4} 
W_1 = \ket{({1/2}^B 1^{AX}) 1/2, m }\bra{1/2, m}= 
C^{(1/2 s)(1 t)}_{(1/2 m)} C^{(1/2 u)(1/2 v )}_{(1 t)} \ket{1/2, u}_A \ket{1/2,s}_B \ket{1/2, v}_X \bra{1/2, m}, \\
\eeq 

where $C^{(**)(**)}_{(**)}$ denote Clebsch-Gordan coupling coefficients and sum over repeated indices is understood. In this expression, $\ket{j,m}$ denote elements of an orthonormal basis for a spin-$j$ representation of $\textrm{SU}(2)$, and $\ket{(j_1 j_2) j,m}$ denote elements of a basis for the irreducible spin-$j$ representation contained in $j_1 \otimes j_2$. From Eqs(\ref{eq:simplestintertwiner1})-(\ref{eq:simplestintertwiner4}), one can verify that $\tr W_1^* V_0=\sqrt{3}$ and $\tr W_1^* V_1=-1$. We can, without loss of optimality, assume that $\lambda_0, \lambda_1, \mu_0$ and $\mu_1$ are real. Using $|\lambda_0|^2+|\lambda_1|^2=1$ and $|\mu_0|^2+|\mu_1|^2=1$. We get 
\beqa\label{eq:sol2111}
\omega^A(1/2,1/2,1/2) & = & 1-\frac{4}{3} \lambda_1^2, \\
\omega^B(1/2,1/2,1/2) & = & 1-\frac{4}{3} \mu_1^2=
1-\frac{4}{3}(\frac{\sqrt{3}}{2} \sqrt{1-\lambda_1^2}-\frac{1}{2} \lambda_1)^2.
\eeqa
The corresponding fidelities are
\beqa\label{eq:fideonetoone}
F^A &=& (1+\omega^A(1/2,1/2,1/2))/2, \\
F^B &=& (1+\omega^B(1/2,1/2,1/2))/2. 
\eeqa 

On Fig.\ref{fidnn1fig}, we have plotted the locus $\{(F^A(\lambda_1),F^B(\lambda_1))\}$. One readily checks that this locus of couples of fidelities correspond with the results in \cite{cerf00:asym}.

%%%%%%%%%%%%%%%%%%%%%%%%%%%%%%%%%%%
\subsection{$n\to n+1$ cloning of qubits}\label{subsection:twoplusone}
%%%%%%%%%%%%%%%%%%%%%%%%%%%%%%%%%%%

We now solve the $2\to 2+1$ case and give the $n\to n+1$ conjectured optimal fidelities. The obtained figures of merit agree with the expected values at the limiting points.

Let us start with the $2\to 2+1$ derivation. First, let us observe that $D(\pi_{1/2}^{\otimes 2})=\{ \pi_0, \pi_1\}$, $D(\pi_{1/2})=\{ \pi_{1/2} \}$, $D_2(\pi_0 \otimes \pi_{1/2})=\{ \pi_{1/2}, \pi_{3/2}\}$, 
$D_2(\pi_1 \otimes \pi_{1/2})=\{ \pi_{1/2}, \pi_{3/2}, \pi_{5/2}\}$. Therefore, according to Sect.\ref{section:generic}, the optimal map we are looking for can be decomposed as a convex sum of five maps as:
\beqa
T= && r(0,1/2,1/2) T(0,1/2,1/2)+r(0,1/2,3/2) T(0,1/2,3/2)+r(1,1/2,1/2) T(1,1/2,1/2)+\\
&& r(1,1/2,3/2) T(1,1/2,3/2)+r(1,1/2,5/2) T(1,1/2,5/2).
\eeqa
Consider the map $T(0,1/2,1/2)$. This map is characterised by an intertwining operator $V$ satisfying 
\beq
V \pi_1= (\pi^A_0 \otimes \pi^B_{1/2} \otimes \pi^X_{1/2}) V.
\eeq
There exists a (Clebsch-Gordan) matrix $C$ such that $\pi^B_{1/2} \otimes \pi^X_{1/2}= C^*(\pi_0^{BX} \oplus \pi_1^{BX})C$. With $V'=CV$, we thus have
\beq
V \pi_1= (\pi_0^A \otimes (\pi_0^{BX} \oplus \pi_1^{BX})) V.
\eeq
From Eq.(\ref{eq:omegacasimir}), one then sees that
\beq
\omega^A(0,1/2,1/2)=\frac{1}{2}(1+\frac{C_2(\pi_0)-C_2(\pi_1)}{C_2(\pi_1)})=0.
\eeq
One can also compute that
\beq
\omega^B(0,1/2,1/2)=1/2.
\eeq
Performing the same analysis for the map $T(0,1/2,3/2)$, one finds that $\omega^A(0,1/2,3/2)=0$ and $\omega^B(0,1/2,3/2)=-1/4$. Since a trivial cp-map that merely prepares clones in a random state achieves $\omega^A=\omega^B=0$, we see that this latter map is useless for cloning. The map $T(1,1/2,5/2)$ also turns out to be useless because $\omega^A(1,1/2,5/2)=-1/2$ and $\omega^B(1,1/2,5/2)=-1/4$. Consider now the map $T(1,1/2,1/2)$. The intertwiner that characterises this map (and that we will again denote $V$) satisfies
\beq
V \pi_1= (\pi^A_1 \otimes \pi^B_{1/2} \otimes \pi^X_{1/2}) V.
\eeq
Again there exists a (Clebsch-Gordan) matrix C such that $V'=CV$ satisifies
\beq\label{eq:intertwin221}
V' \pi_1= (\pi^A_1 \otimes (\pi^{BX}_0 \oplus \pi^{BX}_1)) V'.
\eeq
The space of solutions for Eq.(\ref{eq:intertwin221}) is 2-dimensional: $V'$ is a linear combination of an intertwiner $V'_0$ between $\pi_1$ and $\pi^A_1 \otimes \pi^{BX}_0$, and an intertwiner $V'_1$ between $\pi_1$ and $\pi^A_1 \otimes \pi_1^{BX}$. Accordingly, one finds that
\beqa
\omega^A(1,1/2,1/2) &=& |\lambda(1,1/2,1/2,0)|^2 \frac{1}{2}(1+\frac{C_2(1)-C_2(0)}{C_2(1)})+
|\lambda(1,1/2,1/2,1)|^2 \frac{1}{2}(1+\frac{C_2(1)-C_2(1)}{C_2(1)}) \\
&=&  |\lambda(1,1/2,1/2,0)|^2+\frac{1}{2} |\lambda(1,1/2,1/2,1)|^2,
\eeqa
where $|\lambda(1,1/2,1/2,0)|^2+ |\lambda(1,1/2,1/2,1)|^2=1$ (because $V$ is an isometry). Considering the second set of clones, on finds that
\beq
\omega^B(1,1/2,1/2) = \frac{1}{2}|\mu(1,1/2,1/2,1/2)|^2-\frac{1}{4} |\mu(1,1/2,1/2,3/2)|^2,
\eeq
where $|\mu(1,1/2,1/2,1/2)|^2+ |\mu(1,1/2,1/2,3/2)|^2=1$.
A similar analysis of the map $T(1,1/2,3/2)$ shows that
\beq
\omega^A(1,1/2,3/2)=\frac{1}{2}(|\lambda(1,1/2,3/2,1)|^2-|\lambda(1,1/2,3/2,2)|^2),
\eeq
and 
where $|\lambda(1,1/2,3/2,1)|^2+ |\lambda(1,1/2,3/2,2)|^2=1$, and 
\beq
\omega^B(1,1/2,3/2) = \frac{1}{2}|\mu(1,1/2,3/2,1/2)|^2-\frac{1}{4} |\mu(1,1/2,3/2,3/2)|^2,
\eeq
where $|\mu(1,1/2,3/2,1/2)|^2+|\mu(1,1/2,3/2,3/2)|^2=1$. Extremising numerically, we have found that optimal asymmetric cloning machines are such that $r(1,1/2,1/2)=1$. Adopting lighter notations, the fidelities can be written as 
\beq
F^A = 1-\frac{x^2}{4}, \;
F^B = 1-\frac{(x+\sqrt{2} y)^2}{4},
\eeq
where $x^2+y^2=1$.
Accodring to Sect.\ref{section:generic}, optimal machines should be of the form (Schršdinger picture)
\beq
T_{\textrm{opt}}: \mathcal{B}(\mathcal{H}^{\otimes 2}_+) \to  \mathcal{B}(\mathcal{H}^{\otimes 3}):
\rho^{\otimes n} \to (a^* E_{3/2}+ b^* E_{1/2}) (\rho^{\otimes 2} \otimes \gras{1}) 
(a E_{3/2}+ b E_{1/2}),
\eeq
where $E_{3/2}$ and $E_{1/2}$ are the projectors onto the irreducible subspaces obtained from the decomposition $\pi_1 \otimes \pi_{1} \approx \pi_{3/2} \oplus \pi_{1/2}$. Also, $2(2 a^2+b^2)/3=1$ because $T_{\textrm{opt}}$ has to be trace-preserving.
Note that the fidelities for the limiting
cases are recovered, that is
\begin{eqnarray}
  F_1&=&F_2=1 \quad\Longleftrightarrow\quad F_3=\frac{1}{2} \nonumber\\
  F_1&=&F_2=F_3=F_{\textrm{sym}}(2\to 3)=\frac{11}{12} \nonumber\\
  F_1&=&F_2=F_{\textrm{sym}}(1\to 2)=\frac{5}{6} \quad\Longleftrightarrow\quad
  F_3=1 .
\end{eqnarray}

The previous computation strongly supports that the conjectured machines found by combining linearly the projectors $E_{(n-1)/2}$ and $E_{(n-1)/2}$ coming from the decomposition $\pi_{n/2} \otimes \pi_{1/2} \approx \pi_{(n-1)/2} \oplus \pi_{(n+1)/2}$. The corresponding fidelities are given by
\begin{equation}\label{fidnn1}
    F_1=F_2=\ldots=F_n=1-\frac{2}{n(n+2)} x^2
    \hspace{1 cm} F_{n+1}=1-\frac{1}{2}\left(\sqrt{\frac{n}{n+2}}
    x-y\right)^2,
\end{equation}
with $x^2+y^2=1$.

\begin{figure}[h]
\begin{center}
\epsfig{figure=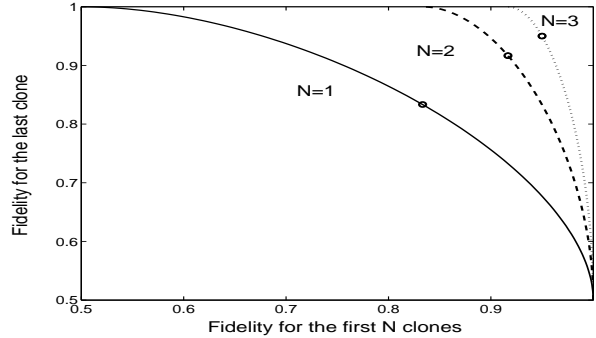,width=80mm,height=45mm}
\caption{Fidelities for the $n\to n+1$ cloning machines.The curves for $n=1,2$ are known to be optimal. For $n>2$  the curve is conjectured to be optimal. Dots on the curves correspond to symmetric machines.} \label{fidnn1fig}
\end{center}
\end{figure}

These fidelities are depicted in figure \ref{fidnn1fig} for
$n=1,\ldots,3$, where only the relevant part of the curve is
shown. The extreme cases are now
\begin{eqnarray}
  F_1&=&F_2=\ldots=F_n=1 \quad\Longleftrightarrow\quad F_{n+1}=\frac{1}{2}
  \quad\Longleftrightarrow\quad y=0 \nonumber\\
  F_1&=&F_2=\ldots=F_{n+1}=F_{S}(n\to n+1)=\frac{n^2+3n+1}{n^2+3n+2}
  \quad\Longleftrightarrow\quad y= \left(\sqrt{\frac{n}{2(n+1)}}\right)\nonumber\\
  F_1&=&F_2=\ldots=F_n=F_{S}(n-1\to n)=\frac{n^2+n-1}{n(n+1)}
  \quad\Longleftrightarrow\quad  F_{n+1}=1
  \quad\Longleftrightarrow\quad y=  \left(\sqrt{\frac{n+2}{2(n+1)}}\right).
\end{eqnarray}
When $n\rightarrow\infty$, an unlimited number of copies of the initial state are available. Then, it is possible to completely determine it and prepare a new identical copy. Then $F_{n+1}\rightarrow 1$ as expected.

%%%%%%%%%%%%%%%%%%%%%%%%%%%%%%%%%%%
\subsection{$1 \to 1+n$ cloning of qubits}\label{section:qubitetmesure}
%%%%%%%%%%%%%%%%%%%%%%%%%%%%%%%%%%%

The method described in Section \ref{section:generic} has been applied to the case of qubits where one set, $A$, consists of one clone, and the other set, $B$, consists of $n$ clones ($n \geq 2$). 

The optimal fidelities are derived in Appendix \ref{section:qubitetmesurecalc} and are given by
\beq\label{eq:onetoneven}
F^A = 1-\frac{2}{3}y^2,\hspace{1cm}
F^B = \frac{1}{2}+\frac{1}{3n}(y^2+\sqrt{n(n+2)}xy),
\eeq
where $x^2+y^2=1$.

One can check that these fidelities correspond to those obtained using the conjectured form of the optimal cloning map.

Let us discuss these expressions. First of all, imposing $F^A=1$ implies $F^B=1/2$. This fact is consistent with the idea that in order to prepare a perfect clone, one has to take it from the input and not let it interact with any system \cite{woot82}. Then, no quantum information is available to prepare the $N$ supplementary clones and the best one can do is to prepare the $N$ qubits in a completely random state, thus achieving a fidelity $1/2$.

Second, if one requires the clones $B$ to have the fidelity of an optimal symmetric $1 \to n$ cloning machine, namely $F^B=\frac{2n+1}{3n}$, one finds that $y=\sqrt{\frac{n+2}{2n+2}}$, which gives $F^A=\frac{2n+1}{3(n+1)}$. Interestingly this fidelity is larger than $1/2$: in order to produce $n$ optimal clones from one input,not all quantum information need be used and some quantum information remains to prepare a "good" $(n+1)$-th clone.

Let us now turn to the case of large $n$. It is well known that there are deep connections between cloning and state estimation \cite{gisi97,keyl02,brus98}. In particular, for universal symmetric cloning, it appears that there is a  correspondence between $n \to \infty$ cloning machines and state estimation devices \cite{brus98:concat}. Such a relation still holds in the asymmetric case. Following the lines of \cite{brus98:concat}, one finds that, in the limit $n \to \infty$, asymmetric $1 \to 1+n$ cloning machines interpolate between (trivial) machines leaving the quantum system unchanged, and a measuring device estimating destructively the input state. In the limit $n \to \infty$, Eqs. (\ref{eq:onetoneven}) become:
\begin{equation}\label{eq:11inf}
F^A = 1-\frac{2}{3} y^2, \qquad F^{\textrm{meas}} =
\frac{1}{2}+\frac{1}{3}y\sqrt{1-y^2},
\end{equation}
where only the case $0 \leq y \leq 1/\sqrt{2}$ should be considered. One readily checks that the two extreme cases are found:(i) When $F^A=1$, one finds $F^{\textrm{meas}} = 1/2 $, which translates the fact that no information can be gained if the input state is unperturbed. (ii) The maximum value of $F^{\textrm{meas}}$ is 2/3, which is consistent with \cite{mass95}. In that case, of course, $F^A=2/3$ too. Between these two cases, the relations (\ref{eq:11inf}) express the trade-off between the acquisition of knowledge about the state of a quantum system and the disturbance undergone by this system. Actually, such a trade-off had been previously studied in \cite{bana01}, in the form of an inequality. So, our machine provides a concrete means to achieve measurements saturating this inequality.

%%%%%%%%%%%%%%%%%%%%%%%%%%%%
\subsection{$1 \to 1+1$ cloning of qudits}\label{section:qudit}
%%%%%%%%%%%%%%%%%%%%%%%%%%%%

$1 \to 1+1$ asymmetric machines were first introduced in \cite{cerf00:asym}. The main interest of such machines is that they are useful in assessing the security of quantum cryptographic protocols \cite{cerf:dlevelcrypto}. However, such machines were known to be optimal only in the case of qubits. In principle, we could apply the method presented in Section \ref{section:generic} to prove the optimality of these. We did not perform such a calculation. Alternatively, one could prove the optimality of these asymmetric machines using the isomorphism between CP maps and positive semidefinite operators \cite{ibli04,fiur05}. Perhaps not surprisingly, one finds that optimal cloning machines \emph{are} of the form (\ref{eq:conjecture}). Under $\pi^{\otimes 2}$, $\mathscr{H}^{\otimes 2}$ decomposes as a
$\frac{d(d+1)}{2}$-dimensional symmetric  subspace
$\mathscr{H}_2^+$ and a $\frac{d(d-1)}{2}$-dimensional
anti-symmetric subspace $\mathscr{H}_2^-$: \beq \pi^{\otimes 2}
\approx \pi_2^+ \oplus \pi_2^-,  \hspace{1cm} \mathscr{H}^{\otimes
2} \approx \mathscr{H}_2^+ \oplus \mathscr{H}_2^-, \hspace{1cm}
\gras{1}^{\otimes 2}=S_2+A_2. \eeq

Let $\{\ket{i}, i=0 \ldots d-1 \}$ denote an orthonormal basis of
$\mathscr{H}=\gras{C}^d$. Clearly, \beq S_2=\frac{1}{2}\sum_{i,j=0}^{d-1}(
\ket{i}\bra{i} \otimes \ket{j}\bra{j}+ \ket{i}\bra{j} \otimes
\ket{j}\bra{i}), \hspace{1cm} A_2=\frac{1}{2}\sum_{i,j=0}^{d-1}(
\ket{i}\bra{i} \otimes \ket{j}\bra{j}- \ket{i}\bra{j} \otimes
\ket{j}\bra{i}). \eeq

According to Eq.(\ref{eq:conjecture}), in Schr\"odinger picture,
the optimal cloning map is of the form 
\beq 
T'_{\textrm{opt}}:
\mathscr{B}(\mathscr{H}) \to \mathscr{B}(\mathscr{H}^{\otimes 2}):
\rho \to (\alpha^* S_2 + \beta^* A_2) (\rho \otimes \gras{1})
(\alpha S_2 + \beta A_2). \eeq Since $T'_{\textrm{opt}}$ should be
trace-preserving, we have \beq \frac{d+1}{2} |\alpha|^2
+\frac{d-1}{2} |\beta|^2=1. 
\eeq

Since the map $T'_{\textrm{opt}}$ is covariant, the fidelity is the
same for all input state, and can be calculated with a particular
state, say $\ket{0}\bra{0}$. Straightforwardly, we get: 
\beqa 
F^A=\tr (\ket{0}\bra{0} \otimes \gras{1})T'_{\textrm{opt}}(\ket{0}\bra{0})=\frac{d+3}{4} |\alpha|^2+\frac{d-1}{4} |\beta|^2+\frac{d-1}{4}(\alpha^* \beta+ \alpha \beta^*). \\
F^B =\tr ( \gras{1} \otimes \ket{0}\bra{0}) T'_{\textrm{opt}}(\ket{0}\bra{0})=\frac{d+3}{4} |\alpha|^2+\frac{d-1}{4} |\beta|^2-\frac{d-1}{4} (\alpha^* \beta+ \alpha \beta^*). \\
\eeqa

Direct calculations show that these machines correspond to the universal asymmetric cloning machines of qudits introduced in \cite{cerf00:asym}.

%%%%%%%%%%%%%%%%%%%%%%%%%%%%%%%%%%%
\subsection{$1 \to 1+1+1$ cloning of qubits}\label{sec:unplusunplusun}
%%%%%%%%%%%%%%%%%%%%%%%%%%%%%%%%%%%

We now turn to asymmetric cloning machines with more than two sets of clones. The simplest example of such a cloning machine is a $1 \to 1+1+1$ cloning machine of qubits which we shall exhibit now. Following Sect.\ref{section:generic}, our first task is to determine the representations $\pi_{\beta}$ satisfying $\pi_{1/2}^{\otimes 3} \otimes \pi_{\beta} \supset \pi_{1/2}$. One directly finds $\pi_{\beta}=\pi_0,\pi_1,\pi_2$. Accordingly, optimal $1 \to 1+1+1$ cloning maps are of the form
\bed
T=r_0 T_0+ r_1 T_1+ r_2 T_2,
\eed
where $T_{\beta}: \mathscr{B}(\mathscr{H}_{1/2}^{\otimes 3}) \to \mathscr{B}(\mathscr{H}_{1/2}):A \to V_{\beta}^* (A \otimes \gras{1}_{\beta}) V_{\beta}$, with $(\pi_{1/2}^{\otimes 3} \otimes \pi_{\beta})V_{\beta}=V_{\beta} \pi_{1/2}$, and where $r_0, r_1, r_2 \geq 0, r_0+r_1+r_2=1$. The fidelities for the three clones are given by 
\bed
F^A=\frac{1}{2}(1+\omega^A)\\
F^B=\frac{1}{2}(1+\omega^B)\\
F^C=\frac{1}{2}(1+\omega^C)\\
\eed
where 
\bed
\omega^A=r_0 \omega^A_0+ r_1 \omega^A_1+ r_2 \omega^A_2,\\
\omega^B=r_0 \omega^B_0+ r_1 \omega^B_1+ r_2 \omega^B_2,\\
\omega^C=r_0 \omega^C_0+ r_1 \omega^C_1+ r_2 \omega^C_2.\\
\eed
Let us work out expressions for $\omega^A_0,\omega^B_0,\omega^C_0$. Let us consider the operator $V_0$. It satisfies 
\bed
(\pi_{1/2}^A \otimes \pi_{1/2}^B \otimes \pi_{1/2}^C \otimes \pi_{0}^X) V_0=V_0 \pi_{1/2}.
\eed
Considering the reduction order $A \otimes B \otimes C \otimes X \to A\otimes BC \otimes X \to A \otimes BCX$, one finds that $V_0$ is of the form $V_0=\lambda_0 E_{00}+\lambda_1 E_{01}$, where $|\lambda_0|^2+|\lambda_1|^2=1$, and where 
\bed
E_{00}=C^{(1/2 u)(0 0)}_{(1/2 m)} C^{(1/2 v)(1/2 w)}_{(00)} \ket{1/2 u}_A \ket{1/2 v}_B \ket{1/2 w}_C \ket{00}_X \bra{1/2 m}
\eed
\bed
E_{01}=C^{(1/2 u)(1 k)}_{(1/2 m)} C^{(1 l)(0 0)}_{(1 k)} C^{(1/2 v)(1/2 w)}_{(1 l)} \ket{1/2 u}_A \ket{1/2 v}_B \ket{1/2 w}_C \ket{1 l}_X \bra{1/2 m}.
\eed
Using Eq.(\ref{eq:omegacasimir}), one finds that
\bed
\omega_0^A=1-\frac{4}{3}\lambda_1^2
\eed
Similarly, considering the reduction order $A \otimes B \otimes C \otimes X \to B \otimes AC \otimes X \to B \otimes ACX$ (resp.$A \otimes B \otimes C \otimes X \to C \otimes BA \otimes X \to C \otimes BAX$), one finds that
\bed
\omega_0^B=1-\frac{4}{3}\mu_1^2,\hspace{1cm}
\omega_0^C=1-\frac{4}{3}\eta_1^2,
\eed
where $|\mu_0|^2+|\mu_1|^2=1$, and where $|\eta_0|^2+|\eta_1|^2=1$. According to 
Eq.(\ref{eq:relationmulambda}), the following relations hold
\begin{equation}
\label{mulambdas1111}
    \mu_1=\frac{\sqrt 3\lambda_0-\lambda_1}{2},\quad\quad\quad\quad
    \eta_1=-\frac{\sqrt 3\lambda_0+\lambda_1}{2}.
\end{equation}

With a similar reasoning for the map $T_1$, one finds that
\begin{equation}
  \omega_1^A=1-\frac{4}{3}(\bar\lambda_1^2+\bar\lambda_{1'}^2), \quad\quad\quad\quad
  \omega_1^B=1-\frac{4}{3}(\bar\mu_1^2+\bar\mu_{1'}^2), \quad\quad\quad\quad
  \omega_1^C=1-\frac{4}{3}(\bar\eta_1^2+\bar\eta_{1'}^2) ,
\end{equation}
where
\beq
|\bar\mu_0|^2 +  |\bar\mu_1|^2+|\bar\mu'_1|^2=1,
\eeq
\begin{eqnarray}
\label{etalambdas1111}
\bar\mu_1 &=& \frac{\bar\lambda_0+\bar\lambda_{1'}}{\sqrt 2}\;
 \bar\mu_{1'} = \frac{-\bar\lambda_0+\sqrt 2\,\bar\lambda_1
 +\bar\lambda_{1'}}{ 2} \nonumber\\
 \bar\eta_1 &=& \frac{\bar\lambda_0-\bar\lambda_{1'}}{\sqrt 2}\;
 \bar\eta_{1'} = \frac{-\bar\lambda_0+\sqrt 2\,\bar\lambda_1
 -\bar\lambda_{1'}}{ 2} .
\end{eqnarray}

One also easily checks that $\omega^A_2=\omega^B_2=\omega^C_2=-1/3$, so that the map $T_2$ is useless for optimal cloning, and we can choose $r_2=0$. In summary, optimal cloning machines are found after maximizing
\begin{eqnarray}
  \omega^A &=& r_0\frac{4\lambda_0^2-1}{3}+
  r_1\frac{4\bar\lambda_0^2-1}{3}, \nonumber\\
  \omega^B &=& r_0\frac{(\lambda_0+\sqrt 3\lambda_1)^2-1}{3}+
  r_1\frac{(\bar\lambda_0+\sqrt 2\,\bar\lambda_1
  -\bar\lambda_{1'})^2-1}{3}, \nonumber\\
  \omega^C &=& r_0\frac{(\lambda_0-\sqrt 3\lambda_1)^2-1}{3}+
  r_1\frac{(\bar\lambda_0+\sqrt 2\,\bar\lambda_1
  +\bar\lambda_{1'})^2-1}{3},
\end{eqnarray}
subject to the normalization, and constraints (\ref{mulambdas1111}),(\ref{etalambdas1111}).

Numerical calculations suggest that the optimal solution corresponds to $r_1=1$ ($r_0=0$). In this case, the optimal $1\to 1+2$ machine given above can be recovered taking $\bar\lambda_{1'}=0$. Interestingly, one finds in that case that some quantum information still remains to produce a non-trivial third clone.

Remark: it is very natural to think of using $1 \to 1+1+1$ cloning machines to perform a simultaneous measurement of the three Pauli operators, measuring each Pauli operator at each output of the cloning machine. However such a measurement will not be optimal, as has already been demonstrated in \cite{dari01:join} using a symmetric $1 \to 3$ cloning machine.

%%%%%%%%%%%%%%%%%%%%%%%%%%%%%%
\section{Optical Implementations}\label{sec:implementations}
%%%%%%%%%%%%%%%%%%%%%%%%%%%%%%

We now turn to the issue of implementing some of the machines presented in the previous sections. We will restrict ourselves to optical implementations where qubits are represented by polarisation states of photons: $N$ identical qubit will be represented by $N$ photons in an identical polarisation mode. In the case of symmetric $N \to M$ cloning, such implementations have already been proposed \cite{SWZ}, and demonstrated experimentally \cite{experiments}. Let us first briefly recall how these $N \to M$ cloning machines work. Let 
\beq\label{eq:prepimpl}
\ket{\psi_{\textrm{in}}} \frac{1}{\sqrt{N!}}(\alpha a_{V,s}^\dagger+\beta a_{H,s}^\dagger)^N \ket{\textrm{vac}},
\eeq
denote the input state to clone, where $\ket{\textrm{vac}}$ denotes the vacuum state. The labels $V,H,s$ stand for vertical, horizontal and signal respectively. Cloning is achieved when a mode prepared in a state (\ref{eq:prepimpl}) impinges a crystal where a parametric down-conversion (PDC) process can occur. The hamiltonian describing this process is of the form:
\begin{equation}\label{PDCham}
 H=\gamma (a_{V,s}^\dagger a_{H,i}^\dagger-a_{H,s}^\dagger a_{V,i}^\dagger) + \textrm{ h.\,c. } ,
 \end{equation}
 where 'i' denotes the idler mode. So the state after the crystal is 
\begin{equation}\label{stim}
 \ket{\psi_s} \propto e^{-iHt}(\alpha a_{V,s}^\dagger+\beta a_{H,s}^\dagger)^N\ket{\textrm{vac}} .
\end{equation}

Looking at those cases where there are $M$ photons in the signal
mode, one can see that the optimal fidelities for the $N\to M$
cloning machine for qubits are obtained. Therefore, the successful
realization of the cloning machine is conditioned on the number of
photons at the output. Note that when $M$ photons are observed in
the signal mode, $N$ of them came from the initial state and $M-N$
were produced at the crystal, which means that there are as well
$M-N$ photons in the idler mode. These photons are usually called
anti-clones. The total number of photons is $2M-N$.

\begin{figure}[h]
\begin{center}
\epsfig{figure=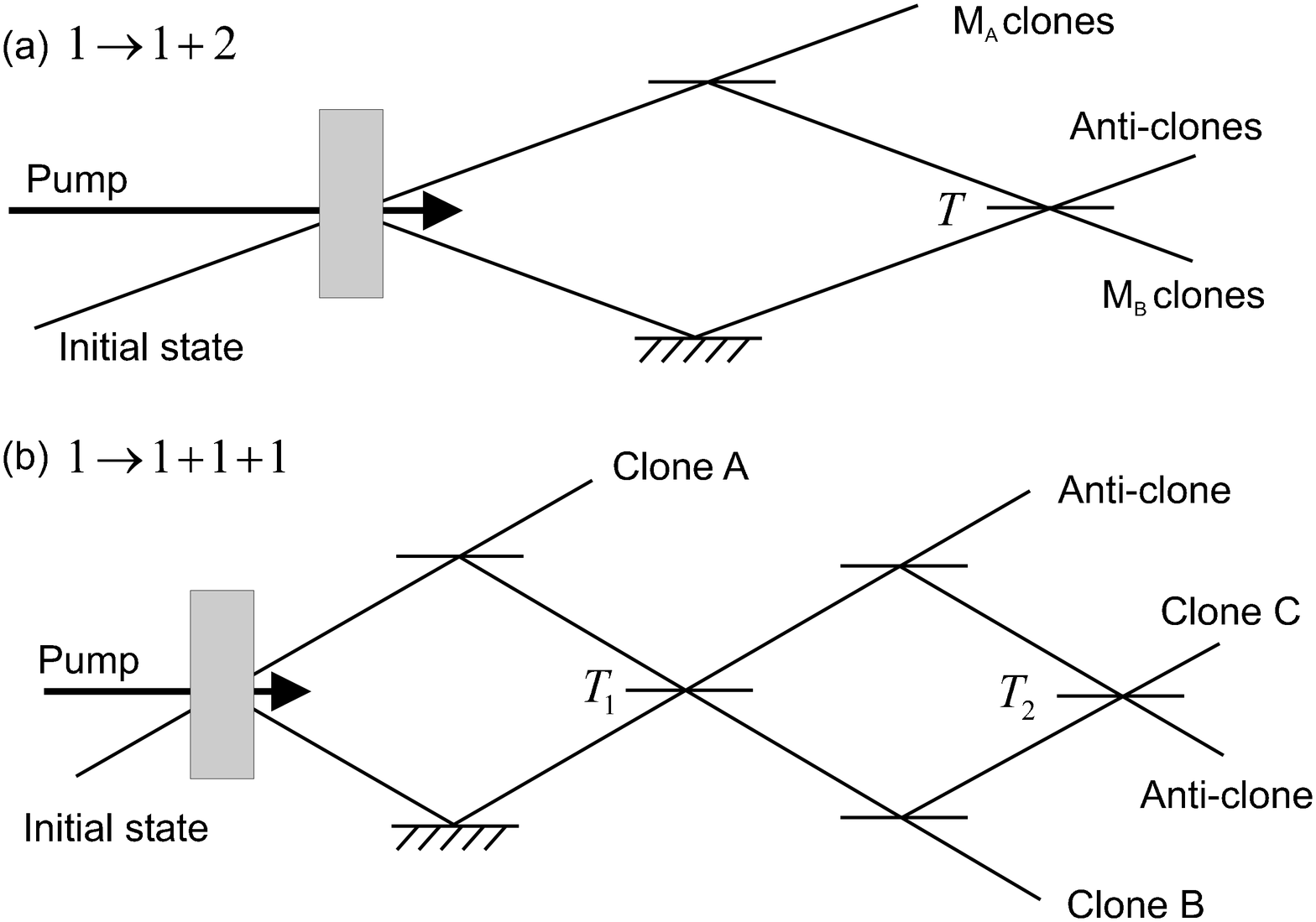,width=90mm,height=60mm}
\caption{Optical implementation of (a) the $1\to 1+2$ 
and (b) the $1\to 1+1+1$ optimal asymmetric cloning machines.}\label{condimpl}
\end{center}
\end{figure}

A modification of this scheme was proposed in \cite{Filip} by Filip, in order to obtain the asymmetric $1\to 1+1$ machine discussed above. His scheme is shown in Fig. \ref{condimpl}a. After a successful $1\to 2$ cloning (that is, when all the three detectors will click), the two clones are split at the first beam splitter, and one of the clones is combined with the anti-clone at a beam-splitter of transmittivity $T$. Long but straightforward algebra shows that the fidelities at the modes 1 and 2 give the $1\to 1+1$ machine, depending on $T$.

\begin{figure}[h]
\begin{center}
\epsfig{figure=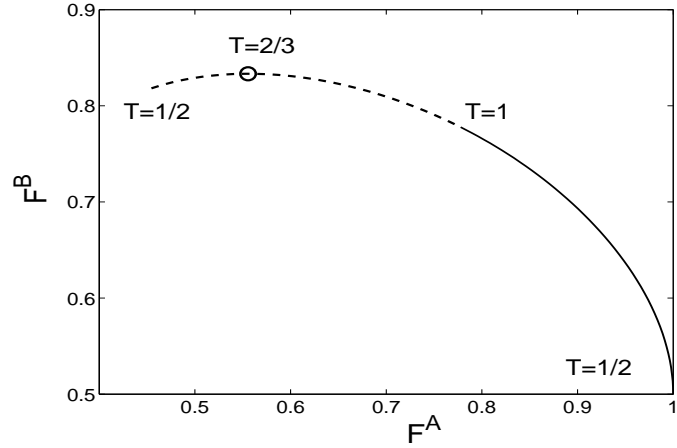,width=90mm,height=60mm}
\caption{Clone fidelities for the optical
  implementation of the $1\to 1+2$ optimal machine. The solid, resp.
  dashed, line represents the case when $M_A=1$ and $M_B=2$, resp.
  $M_A=2$ and $M_B=1$.}\label{figexpfid112}
\end{center}
\end{figure}

The natural question is now whether this modification also
provides the optimal solution for the more general case $N\to
M_A+M_B$. Note that the scheme of \cite{SWZ} gives all $N\to M$
machines, simply conditioning on different number of photons at
the input and output signal mode. We denote by $M_i$ the number of
photons in mode $i$. As said above, the results of \cite{Filip}
imply that conditioned on the fact that there is one photon at the
input signal mode, $N=1$, and one photon at each output mode,
$M_A=M_B=M_a=1$, the optimal $1\to 1+1$ machine is obtained. It is
also easy to see that this machine is covariant, so all the
calculations can be done taking as initial state
$(a_{V,s}^\dagger)^N\ket{\textrm{vac}}$. Then, the state at the output of the
crystal when the total number of photons is $2M-N$ reads
\begin{equation}
\label{pdcout}
    \ket{\psi_s^M}=\sum_{j=0}^{M-N}(-1)^j \sqrt{\binom{M-j}{N}}
    \ket{M-j}_{V,s}\,\ket{j}_{H,s}\,\ket{j}_{V,i}\,\ket{M-N-j}_{H,i} .
\end{equation}
The simplest generalization of the $1\to 1+1$ result corresponds
to $N=1$, $M_A=1$ $M_B=2$. One can check that the evolution of the state (\ref{pdcout}) through the
beam-splitters, where $N=1$ and $M=M_A+M_B=3$, gives the following
fidelities, depending on $T$,
\begin{equation}\label{expfid1121}
    F_1=\frac{4T^2-4T+7}{12T^2-12T+9} \hspace{2cm}
    F_2=F_3=\frac{8T^2-4T+3}{12T^2-12T+9} .
\end{equation}
These fidelities are shown in figure \ref{figexpfid112}. Only the
relevant part for $1/2\leq T\leq 1$ is depicted. Note that for
$T=1$ the optimal symmetric machine is recovered, as expected. If
the transmittivity decreases, the quality of the first clone
increases, while the quality of the two clones in mode 2 worsens.
When $T=1/2$ all the information on mode 2 (and 3) is lost, and a
perfect copy of the initial state is obtained at mode 1.

How can the missing values be obtained? Note that in the previous
expressions, the fidelity for the clone in the first mode is always
larger than that of the two post-selected photons in mode 2. This
suggests a way to find the remaining part of the curve of optimal
fidelities: one has to reverse the post-selection of photons, that
is look at the cases where $M_A=2$ and $M_B=1$. In this way, one
expects to reproduce the situation where the fidelity for the two
clones, now in mode 1, is larger than the fidelity for the single
clone, now in mode 2. Repeating the calculations, but now for
$M_A=2$ and $M_B=1$, one has
\begin{equation}\label{expfid1122}
    F_1=\frac{7T^2-4T+4}{9T^2-12T+12} \hspace{2cm}
    F_2=F_3=\frac{3T^2-4T+8}{9T^2-12T+12} .
\end{equation}
The corresponding curve is also shown in figure
\ref{figexpfid112}. In this way the optimal $1\to 1+2$ case is
completely recovered. Indeed, when $t=2/3$ it is found that the
two photons in mode 1 have fidelities $5/6$, while the photon in
mode 2 has fidelity $5/9$, as it should be. Remarkably, the
fidelities of Eq. (\ref{expfid1121}) are the same as in Eq.
(\ref{expfid1122}), changing $T$ into $1/T$.

Using the same ideas, we also analyzed the case $2\to 2+1$. For
$M_A=2$ and $M_B=1$ one finds, putting $N=2$ in (\ref{pdcout})
\begin{equation}\label{expfid2211}
    F_1=F_2=1-\frac{(2T-1)^2}{4(4T^2-4T+3)} \hspace{2cm}
    F_3=1-\frac{(2T-3)^2}{4(4T^2-4T+3)} ,
\end{equation}
while for $M_A=1$ and $M_B=2$ one has
\begin{equation}\label{expfid2212}
    F_1=F_2=1-\frac{(T-2)^2}{4(3T^2-4T+4)} \hspace{2cm}
    F_3=1-\frac{(3T-2)^2}{4(3T^2-4T+4)} .
\end{equation}
Note that, again, (\ref{expfid2212}) can be obtained from
(\ref{expfid2211}) if $T$ is replaced by $1/T$. These fidelities
are depicted in Fig.\ref{figexpfid221}, they indeed correspond
to the optimal solution given above.

Actually, the case $N\to N+1$ can also be
computed. The obtained fidelities when $M_A=N$ and $M_B=1$ are
\begin{equation}\label{expfidnn1}
    F_1=F_2=1-\frac{(2T-1)^2}{(N+2)(2NT^2-2NT+N+1)} \hspace{2cm}
    F_3=1-\frac{(NT-N+1)^2}{(N+2)(2NT^2-2NT+N+1)} ,
\end{equation}
while the expressions for $M_A=1$ and $M_B=N$ are again given from
these quantities after replacing $T$ by $1/T$. 
One can check that the obtained fidelities  are identical to Eqs. (\ref{fidnn1}), conjectured to
be optimal.

All the previous results give support to the conjecture that all
the $N\to M_A+M_B$ cloning machines are included in
Filip's scheme, as it happened for the symmetric case \cite{SWZ}.
Unfortunately, this is not the case. Indeed, we've checked that
this scheme does not provide the optimal solution when $N=2$,
$M_A=2$ and $M_B=2$. Therefore, we conjecture that this
modification of the symmetric cloning machine implementation only
works for the cases $1\to N+1$ and $N\to N+1$, that is when only
two irreducible representations appear in the conjectured optimal
solution of (\ref{eq:conjecture}).

This optical scheme can also be adapted to realize the optimal 
$1\to 1+1+1$ cloning transformation, see Fig.\ref{condimpl}(b).
Here, the output of a symmetric $1\to 3$ machine
is made asymmetric by combining some of the clones with
anti-clones at two beam splitters, with transmittance $T_1$ and
$T_2$. The obtained fidelities $F^A\geq F^B\geq F^C$,
depending on $T_1$ and $T_2$, are
optimal. The three fidelities are equal when $T_1=T_2=1$.
Other interesting limiting cases are $(F^A,F^B,F^C)=(1,1/2,1/2)$
when $T_1=1/2$, while taking $T_2=1$ gives Eqs.~(\ref{expfid1121})
for the $1\to 1+2$ case. This construction can
easily be generalized further. 
\par

%%%%%%%%%%%%%
\section{Conclusions}
%%%%%%%%%%%%%

In summary, we have introduced a new class of quantum cloning machines, which, helps to get a better understanding of how quantum information can be distributed unequally between several quantum systems. Mutlipartite asymmetric cloning machines have at least two interesting applications. First, some $N \to M_A+M_B$ machines have been proven to be a useful tool in assessing the security of some quantum cryptographic protocols \cite{curt03,acin04,nied05} . Considering $n \to n+1$ machines, we have seen that if one wants to produce $n$ clones from an input with a fidelity which is as high as possible, some  quantum information still remains to produce a non-trivial $n+1$-th clone.  Also, we have seen how $1 \to 1+n$ cloning allow, in the limit of large values of $n$, to study the trade-off between the gain of knowledge about the state of a quantum system and the disturbance undergone by this system. We have also demonstrated feasible optical implementations of some machines. We have seen that the impossibility of perfect cloning translates in the spontaneous emission that unavoidably accompanies stimulated emission. 

Several questions remain open. We here list a few of them. First, it would be very desirable to prove the conjecture about the structure of optimal cloning maps (to disprove it would turn even more interesting). Second, it would certainly be interesting to find cloning machines optimal with respect to the global fidelities as defined by Eq.(\ref{eq:defglobalfid}), instead of single-copy fidelities as we did in this work. Do optimal machines coincide for both figures of merit as in the case of symmetric cloning \cite{keyl98}? 
Another interesting problem is to find closed formulas for optimal cloning. We have the feeling that this problem will not be solvable with the techniques presented here. The reason is that our optimisation requires the computation of a so-called Racah coefficient, for which, to our knowledge, no closed formula exists in general. Many interesting questions regarding implementations also remain. It is tempting to believe that all $N \to M_A+M_B+\ldots$ cloning machines are only limited by spontaneous emission, and can therefore be implemented by splitting clones and anticlones produced by stimulated emission using beam splitters. This question deserves further investigation. Finally, it would be interesting to perform an (optical) experiment demonstrating the concepts analysed here.

\section{Acknowledgements}

We thank N. Cerf, J. Fiurasek, R. Filip, S. Massar and V. Scarani for discussions. S.I. and N.G. acknowledge financial support by the EU project RESQ and the Swiss NCCR. AA acknowledges support from the Spanish MCYT under "Ram\' on y Cajal" grant.

\appendix

%%%%%%%%%%%%%%%%%%%%%%%%%%%%%%%%%%%%%%%%%%%%%
\section{The symmetries of optimal cloning maps}\label{section:symoptimalmap}
%%%%%%%%%%%%%%%%%%%%%%%%%%%%%%%%%%%%%%%%%%%%%

We now use the method of  Keyl-Werner and exploit symmetries in order to characterize the sought optimal cp-maps $T$ \cite{keyl98}. In a sense this subsection can be considered as a summary of their method. There is however a "twist" with respect to their analysis, due to the fact that there are now two sets of clones with different fidelities.

We will work in Heisenberg picture, where states are left unchanged and operators are transformed. For our purpose, it is convenient to represent the cp-map we look for with the Stinespring dilation theorem \cite{keyl02}. This theorem states that any cloning map can be written as
\beq\label{eq:stinespring} 
T:\mathscr{B}(\mathscr{H}^{\otimes M}) \to \mathscr{B}(\mathscr{H}^{\otimes N}_{+}): \mathscr{O} \to
V^*(\mathscr{O} \otimes \textbf{1}_{\mathscr{K}})V, 
\eeq 
where $\textbf{1}_{\mathscr{K}}$ denotes the identity over some auxiliary Hilbert space, $\mathscr{K}$, and where $V:\mathscr{H}^{\otimes N}_{+} \to \mathscr{H}^{\otimes M}\otimes
\mathscr{K}$  is an isometry ($||V \psi||=||\psi||$). 

The figure of merit (\ref{eq:deflocalfid}) we have chosen is such that  the optimal map can of course be sought amongst covariant maps: that is maps $T$ such that $\forall \mathscr{O} \in \mathscr{B}(\mathscr{H}^{\otimes M})$ and $\forall u \in U(d)$, 
\beq 
\pi_N^+(u) \; T(\mathscr{O})\; \pi_N^+(u)^*=T(\pi^{\otimes M}(u) \; \mathscr{O}  \; \pi^{\otimes M}(u)^*). 
\eeq

Indeed, since $\forall u \in U(d)$, the 'translated' map $T_{u}: \mathscr{O} \to \pi_N^+(u)^* \; T(\pi^{\otimes M}(u) \; \mathscr{O} \; \pi^{\otimes M}(u)^*) \; \pi_N^+(u)$ achieves the
same fidelity as $T$, $F^A(T)=F^A(T_u)$. Thus,
\bed 
F^A(T)=\int du F^A(T)=\int du F^A(T_u) \leq F^A(\int du T_u), 
\eed
where $du$ denotes the Haar measure over $U(d)$ \cite{zhelobenko}($\int du=1$). Similarly, $F^B(T) \leq F^B(\int du T_u)$. This proves that if we find an optimal cp-map $T^{\textrm{opt}}$, then $\int du T^{\textrm{opt}}_u$ is covariant and is optimal too. Thus, searching for an optimal map, we can restrict ourselves to $U(d)$-covariant cp-maps.

This covariance property is the first symmetry property we will use: it merely states that, for the figure of merit we have chosen, no state should be preferred by an optimal cloning machine. Covariance simplifies much the analysis because it allows us to use the covariant form of the Stinespring dilation theorem \cite{keyl02}:
\newtheorem{covariantstinespring}{Theorem}
\begin{covariantstinespring}
If $T:\mathscr{B(H}_1) \to \mathscr{B(H}_2)$ denotes a cp-map
covariant with respect to the representations $\{U(d) \ni u \to
\pi_1(u) \in \mathscr{B(H}_1) \}$ and $\{U(d) \ni u \to \pi_2(u)
\in \mathscr{B(H}_2) \}$, then T is of the form
(\ref{eq:stinespring}) with the auxillary space $\mathscr{K}$
being the carrier space of some representation $\pi_{\mathscr{K}}$
of $U(d)$ and V being an "intertwining" operator: \beq (\pi_1(u)
\otimes \pi_{\mathscr{K}}(u))V=V \pi_2(u) \hspace{1cm} \forall
u\in U(d). \eeq
\end{covariantstinespring}

The second symmetry property to use is of course permutation invariance. Consider constructing a cloning machine as follows. We apply the cloning machine described by the cp-map $T$ to our input, and then apply some permutation on the output clones of the set $A$, and some other permutation on the output clones of the set $B$. Clearly, we expect that the performance of the obtained cloning machine will be the same as that of the cloning machine we started from, whatever the applied permutations are. Let us formalise this property. Let $\textrm{Sym}(M_A)$ denote the group of permutations of $M_A$ objects and let $\textrm{Sym}(M_A) \ni p \to \delta(p)$ denote a representation of $\textrm{Sym}(M_A)$ acting on $\mathscr{H}^{\otimes M_A}$ as: 
\bed
\delta(p):\mathscr{H}^{\otimes M_A} \to \mathscr{H}^{\otimes M_A}:
\phi_1 \otimes \ldots \otimes \phi_{M_A} \to \phi_{p(1)} \otimes
\ldots \otimes \phi_{p(M_A)}. 
\eed 
For all admissible cloning map $T$, the permuted map 
\beq
T_p:\mathscr{O} \to T[(\delta^{-1}(p)\otimes\textbf{1}^{\otimes M_B}) \mathscr{O} (\delta(p) \otimes \textbf{1}^{\otimes M_B})]
\eeq
is a cp-map in its own right and yields clones with the same fidelity as $T$:$F^A(T)=F^A(T_p)$, $\forall p\in S_{M_A}$. The fact that 
\bed 
F^A(T)=\frac{1}{M_A!} \sum_{p \in S_{M_A}} F^A(T_p) \leq F^A(\frac{1}{M_A!} \sum_{p \in S_{M_A}} T_p) 
\eed 
implies that searching for an optimal map, we can focus on $\textrm{Sym}(M_A)$-invariant cp-maps. Similarly, considering the second set of clones, we see that we can focus on $\textrm{Sym}(M_B)$-invariant cp-maps.

Consider now the restricted map, $T^A$, obtained upon tracing $T$ over the second set of clones, i.e. 
\bed
T^A:\mathscr{B}(\mathscr{H}^{\otimes M_A}) \to \mathscr{B}(\mathscr{H}^{\otimes N}_+):\mathscr{O} \to T(\mathscr{O} \otimes \gras{1}^{\otimes M_B}).
\eed
Clearly, $T^A$ is a $U(d)$-covariant, $\textrm{Sym}(M_A)$-invariant map, with range in $\mathscr{B}(\mathscr{H}^{\otimes N}_+)$.  One can prove that such a map is \emph{non-degenerate} \cite{keyl98}, that is there exists a constant $\omega^A(T)$ such that 
\beq 
T^A(\partial\pi^{\otimes M_A}(X))=\omega^A(T) \partial\pi_N^+(X),
\eeq
for all $X \in \textrm{su}(d)$. Non-degeneracy of $T^A$ is a manifestation of the "isotropic" nature of  the cloning machine. Indeed the fidelity for the clones A reads 
\beqa 
F^A(T) &= & \frac{1}{d}+\frac{1}{M_A}\textrm{min}_{\psi} \bra{\psi^{\otimes N}} T^A(\partial \pi^{\otimes M_A}
(\ket{\psi}\bra{\psi}-\frac{\gras{1}}{d})) \ket{\psi^{\otimes N}}
\nonumber \\
& = & \frac{1}{d}+\frac{\omega^A(T)}{M_A} \textrm{min}_{\psi} 
\bra{\psi^{\otimes N}} T^A(\partial \pi_N^+(\ket{\psi}\bra{\psi}-
\frac{\gras{1}}{d})) \ket{\psi^{\otimes N}} \nonumber \\
& = & \frac{1}{d}(1+\frac{N}{M_A} \omega^A(T) (d-1)) 
\eeqa 
The quantity $\frac{N}{M_A} \omega^A(T)$ can be interpreted as the so-called shrinking factor \cite{brus98:concat}. Clearly, the restricted map $T^B$ associated with the second set of clones is endowed with the same properties as $T^A$ and its cloning quality can also be characterised by some shrinking factor $N\omega^B(T)/M_B$. It is because of non-degeneracy that we said in Sect.\ref{section:algebra} that the clones of each set are charcterised by a single quantity.

We now show permutation invariance allows to decompose the sought cp-map as a convex combination of simpler cp-maps. Let
\beq 
\pi^{\otimes M_A}= \oplus_{\alpha} \pi_{\alpha}, \hspace{1cm} \mathscr{H}^{\otimes M_A}= \oplus_{\alpha} \mathscr{H}_{\alpha},
\hspace{1cm}\gras{1}^{\otimes M_A}=\sum_{\alpha} E_{\alpha} 
\eeq
denote the decomposition theory of $\pi^{\otimes M_A}$. $E_{\alpha}$ is the projector onto $\mathscr{H}_{\alpha}$. Clearly $[E_{\alpha},\pi^{\otimes M_A}(u)]=0 \; \forall u \in \textrm{U}(d)$. Hence, by Shur's lemma, $T(E_{\alpha} \otimes \gras{1}^{\otimes M_B})$ is proportional to the projector onto $\mathscr{H}^+_N$: $T(E_{\alpha} \otimes \gras{1}^{\otimes
M_B})=r_{\alpha} S_N$. So 
\bed 
T_{\alpha}:\mathscr{B(H}_{\alpha}) \to \mathscr{B(H}_N^+): \mathscr{O} \to r_{\alpha}^{-1} T(E_{\alpha} \otimes \gras{1}^{\otimes M_B} \mathscr{O} E_{\alpha} \otimes \gras{1}^{\otimes M_B}) 
\eed 
is a unital cp-map.

Also, the commutant of (the algebra linearly generated by) all unitaries $\pi^{\otimes M_A}(g)$ is the algebra linearly generated by permutation operators $\delta(p), p \in \textrm{Sym}(M_A)$.

Clearly, $\forall p \in \textrm{Sym}(M_A), \forall \mathscr{O} \in \mathscr{B}(\mathscr{H}^{\otimes M_A})$, we have 
\bed
T[\mathscr{O} (\delta(p) \otimes \gras{1}^{\otimes M_B})]=T[(\delta(p) \otimes \gras{1}^{\otimes M_B}) \mathscr{O}].
\eed
Now since each $E_{\alpha}$ is a linear combination of permutation operators and since $E_{\alpha}^2=E_{\alpha}$, we have that $T[\mathscr{O} (E_{\alpha} \otimes \gras{1}^{\otimes M_B})]=T[(E_{\alpha} \otimes \gras{1}^{\otimes M_B}) \mathscr{O} (E_{\alpha} \otimes \gras{1}^{\otimes M_B})]$. Hence the following decomposition holds
\bed 
T(\mathscr{O})= \sum_{\alpha} r_{\alpha}
T_{\alpha}(\mathscr{O}). 
\eed 
Similarly, we can decompose each $T_{\alpha}$ according to the irreducible representations contained in $\pi^{\otimes M_B}$. Thus, we get 
\beq
T=\sum_{\alpha_1 \in D(M_A)} \sum_{\alpha_2 \in D(M_B)}
r_{\alpha_1,\alpha_2} T_{\alpha_1,\alpha_2}, 
\eeq 
where $D(M_A)=\{\alpha_1: \pi_{\alpha_1} \subset \pi^{\otimes M_A} \}$, $D(M_B)=\{\alpha_2: \pi_{\alpha_2}  \subset \pi^{\otimes M_B} \}$ and the coefficients $r_{\alpha_1,\alpha_2}$ are positive constants summing to unity.

It remains to decompose each map $T_{\alpha_1,\alpha_2}$ using the \emph{covariant form} of the Stinespring theorem. The convex decomposition of $T_{\alpha_1,\alpha_2}$ is the same as the reduction theory of $\pi_{\mathscr{K}}$ into irreducibles. Let $\gras{1}_{\mathscr{K}}=\sum_{\beta} \gras{1}_{\beta}$ denote this reduction. $V^*(\gras{1}_{\alpha_1} \otimes \gras{1}_{\alpha_2} \otimes \gras{1}_{\beta})V=r_{\beta} \gras{1} \Rightarrow T_{\alpha_1,\alpha_2,\beta}: \mathscr{O} \to r_{\beta}^{-1} V(\mathscr{O} \otimes \gras{1}_{\beta})V^*$ is a unital cp-map and the decomposition \beq T_{\alpha_1,\alpha_2}=\sum_{\beta}
r_{\beta} T_{\alpha_1,\alpha_2,\beta} \eeq holds. In turn, this decomposition induces the decompositions 
\beqa
\omega^A &=& \sum_{\alpha_1,\alpha_2} r_{\alpha_1,\alpha_2}
\sum_{\beta} r_{\beta} \omega^A_{\alpha_1,\alpha_2,\beta} \\
\omega^B &=& \sum_{\alpha_1,\alpha_2} r_{\alpha_1,\alpha_2}
\sum_{\beta} r_{\beta} \omega^B_{\alpha_1,\alpha_2,\beta} \\
\eeqa

\textbf{Relation between the fidelities of the clones}. We will now characterise the intertwining operator $V$ and see how $\omega^A$ and $\omega^B$ are related. Addressing the first problem requires that we take care of the \emph{order} in which the decomposition of a representation into irreducible components is carried out (this will be clarified below). Addressing the second problem requires that we can connect these orders of decomposition with each other.

Consider a single map $T_{\alpha_1,\alpha_2,\beta}$, and let us solve the equation
\beq
V \pi_N^+(g)=(\pi^A_{\alpha_1}(g) \otimes \pi^B_{\alpha_2}(g) \otimes \pi^X_{\beta}(g)) V
\eeq

We will consider two manners to reduce $\pi^A_{\alpha_1} \otimes \pi^B_{\alpha_2} \otimes \pi^X_{\beta}$. The first manner first reduces the representation $\pi_{\alpha_2}$ (associated with the second set of clones, $B$), with the representation $\pi_{\beta}$ (associated to the auxiliary system, $X$), and then the resulting representation $\pi_{\alpha_1}$ (associated with the first set of clones, $A$): 
\beq\label{eq:reducorder1} 
A \otimes B \otimes X \to A \otimes BX \to ABX.
\eeq 
The second manner is: 
\beq\label{eq:reducorder2}
A \otimes B \otimes X \to B \otimes AX \to BAX.
\eeq

Let us  consider the first reduction order. Then
\bed
\pi_{\alpha_1} \otimes \{ \pi_{\alpha_2} \otimes \pi_{\beta} \}
\approx \sum_a \bigoplus_{i_a=1}^{m_a} \pi^+_{N,i_a} \oplus
\pi_{\textrm{rest}}, 
\eed 
where $m_a$ denotes the multiplicity of $\pi_N^+$ in $\pi_{\alpha_1} \otimes \pi_a$, and where $\pi_{\textrm{rest}}$ contains no copy of $\pi_N^+$.  Let us suppose that $m_a \leq 1 \; \forall a$. The general case is not more complicated to treat, but this assumption will allow us to adopt lighter notations, and at least it holds in all cases exhibited in this paper. Then, we can rewrite the last equivalence as:
\beq 
\pi_{\alpha_1} \otimes \{ \pi_{\alpha_2} \otimes \pi_{\beta}\} \approx  \bigoplus_{a} \pi^+_{N,a} \oplus \pi_{\textrm{rest}},
\eeq 
and (up to unitaries) $V$ satisfies 
\beq 
(\bigoplus_{a} \pi^+_{N,a}(g) \oplus \pi_{\textrm{rest}}(g))V=V \pi^+_N(g). 
\eeq
From this relation and Shur's lemma,  we infer that there exist coefficients $\lambda_a$ such that $V=\sum_a \lambda_a V_a$, where $V_a:
\mathscr{H}^{\otimes N}_+ \to \mathscr{H}^{\otimes N}_{+,a}$ is the unique intertwiner between $\pi_N^+$ and $\pi_{N,a}^+$. One can verify the following properties:
\beq\label{eq:orthintertwiners} 
V_a^* V_b= \delta_{ab} S_N, 
\eeq
\bed 
T_{\alpha_1,\alpha_2,\beta}  \; \textrm{is unital} \Rightarrow  \sum_a |\lambda_a|^2=1,
\eed
where $S_N$ denotes the identity over $\mathscr{H}_N^+$.

Non-degeneracy of $T_{\alpha_1,\alpha_2,\beta}$ is now expressed as 
\beq \omega^A_{\alpha_1,\alpha_2,\beta} \partial\pi_{N}^+(X)=
\sum_a  |\lambda_a|^2  V_a^* (\partial\pi_{\alpha_1}(X) \otimes
\gras{1}_{a}) V_a,    \; X \in \textrm{su}(d). 
\eeq 
At this point, it is possible to express $\omega^A$ as a function of Casimir numbers $C_2(\pi_a)$, as in \cite{keyl98}. We get
\beq\label{eq:omegacloneA}
\omega^A_{\alpha_1,\alpha_2,\beta}=\frac{1}{2} +\frac{1}{2
C_2(\pi_N^+)} \{C_2(\pi_{\alpha_1})-\sum_a |\lambda_a|^2
C_2(\pi_a)\}. 
\eeq

In this expression, the sum runs over all $a \in D_N(\pi_{\alpha_2} \otimes \pi_{\beta})=\{a: \pi_N^+ \subset \pi_{\alpha_2} \otimes \pi_{\beta} \otimes \pi_a \}$. In the case of qubits (SU$(2)$), where irreducible representations are labelled by positive half-integer
numbers $j$, we have $C_2(j)=j(j+1)$. Explicit expressions of
$C_2(\pi_a)$ for irreducible representations of SU$(d)$ can be
found in \cite{zhelobenko,keyl98}.

What about the second set of clones? Eq.(\ref{eq:omegacloneA}) has been derived following the reduction order (\ref{eq:reducorder1}), and the fact that $V=\sum_a \lambda_a V_a$. If instead, we had used the reduction order (\ref{eq:reducorder2}), we would have found that 
\bed
V=\sum_b \mu_b W_b,
\eed
where $\pi_{\alpha_1} \otimes \pi_{\beta} \approx \oplus_b \pi_b$, and where $W_b$ intertwines $\pi^+_N$ and the (unique) copy of $\pi_N^+$ contained in $\pi_{\alpha_2} \otimes \pi_b$ (when any). Thus we get
\beq\label{eq:omegacloneB}
\omega^B_{\alpha_1,\alpha_2,\beta}=\frac{1}{2}+\frac{1}{2C_2(\pi_N^+)}
\{C_2(\pi_{\alpha_2})-\sum_b |\mu_b|^2 C_2(\pi_b)\}. 
\eeq 

In this expression, the sum runs over all $b \in D_N(\pi_{\alpha_1} \otimes \pi_{\beta})$. All we need now, in order to quantify the trade-off between the qualities of the two sets of clones, is a relation between the coefficients $\lambda_a$ and the coefficients $\mu_b$. It is easy to find such a relation: just observe that $V=\sum_a \lambda_a V_a=\sum_z \mu_z W_z \Rightarrow \sum_a \lambda_a W^*_{b} V_a=\sum_z \mu_z W^*_{b} W_z =\mu_{b} S_N$. Thus 
\beq\label{eq:intertwinchgebasis}
\mu_b=\frac{1}{\textrm{dim}\mathscr{H}^{\otimes N}_+}\sum_a
\lambda_a  \tr W^*_{b} V_a. 
\eeq

N.B. The quantity $\tr W^*_b V_a$ is known in representation theory as the Racah coefficient. 

%%%%%%%%%%%%%%%%%%%%%%%%%%%%%%%%%%%%%%%%%%%%%%
\section{calculations related to $1 \to 1+n$ cloning of qubits}\label{section:qubitetmesurecalc}
%%%%%%%%%%%%%%%%%%%%%%%%%%%%%%%%%%%%%%%%%%%%%%

We are looking for a map $T:\mathscr{B}(\mathscr{H} \otimes \mathscr{H}^{\otimes n}) \to \mathscr{B}(\mathscr{H})$. According to Section \ref{section:generic}, $T$ decomposes as 
\bed
T=\sum_{\alpha_2 \in D(\pi^{\otimes n})} \sum_{\beta \in
D_1(\pi_{1/2} \otimes \pi_{\alpha_2})} r(\alpha_2,\beta)
T(\alpha_2,\beta). 
\eed

$D(\pi^{\otimes n})$ is given by the decomposition theory of $\pi_{1/2}^{\otimes n}$, which is well-known \cite{zhelobenko}:
\bed 
\pi_{1/2}^{\otimes n}=\sum_{s \in I_n}
\bigoplus_{i_s=1}^{m_s} \pi_{s,i_s} 
\eed 
In this expression, $m_s$ denotes the multiplicity of the representation $\pi_s$, and $I_n=\{1/2,\ldots, n/2\}$ when $n$ is odd, and $I_n=\{0, \ldots, n/2\}$ when
$n$ is even. Accordingly, 
\bed T=\sum_{s \in I_n}
\sum_{i_s=1}^{m_s} r(s,i_s) T(s,i_s), 
\eed
where $T(s,i_s): \mathscr{B}(\mathscr{H}_{1/2} \otimes \mathscr{H}_{s,i_s}) \to \mathscr{B}(\mathscr{H}_{1/2})$, $r(s,i_s) \geq 0$ and $\sum_{s,i_s} r(s,i_s)=1$. Before decomposing the map $T$ any further, we remark that, for fixed $s$, all maps $T(s,i_s)$ are isomorphic.  Therefore, as far as optimality is concerned, families of cloning machines with the same values of $r_s=\sum_{i_s} r(s,i_s)$ are equivalent. This fact allows to get rid of the multiplicities of each $s$ and simply write
\beq
T=\sum_{s \in I_n}  \sum_{\beta \in D_1(\pi_{1/2} \otimes \pi_s)} r(s,\beta) T(s,\beta).
\eeq

Let us characterise $D_1(\pi_{1/2} \otimes \pi_s)$. We have
$\pi_{1/2} \subset \pi_{1/2} \otimes \pi_s \otimes \pi_{\beta}
\iff \pi_{\beta} \subset \pi_{1/2}^{\otimes 2} \otimes \pi_s$.
There are three cases to consider: Case A: $s=0$, which yields
$D_1(\pi_{1/2} \otimes \pi_0)=\{0,1\}$. This case occurs whenever $n$
is even. Case B: $s=1/2$; which yields $D_1(\pi_{1/2} \otimes
\pi_{1/2})=\{1/2,3/2\}$. This case occurs whenever $n$ is odd.  Case
C: $s > 1/2$, which yields $D_1(\pi_{1/2} \otimes
\pi_s)=\{s-1,s,s+1\}$. This case occurs whenever $n > 1$.

Let us start with case A. So suppose that in the convex decomposition of the
cloning map, $T$, some map $T(0):
\mathscr{B}(\mathscr{H}_{1/2} \otimes \mathscr{H}_0) \to
\mathscr{B}(\mathscr{H}_{1/2})$ appears. $T(0)$ decomposes as $T(0)=r(0,0)
T(0,0)+ r(0,1) T(0,1)$. $T(0,0)$ and $T(0,1)$ have the following
structure:
 \beqa
T(0,0) &:& \mathscr{O} \to V(0,0)^*(\mathscr{O} \otimes \gras{1}_0) V(0,0), \\
T(0,1) &:& \mathscr{O} \to V(0,1)^*(\mathscr{O} \otimes \gras{1}_1) V(0,1),
 \eeqa
 where $(\pi_{1/2}^A \otimes \pi_0^B \otimes \pi_0^X) V(0,0)= V(0,0) \pi_{1/2}$, and
 $(\pi_{1/2}^A \otimes \pi_0^B \otimes \pi_1^X) V(0,1)= V(0,1) \pi_{1/2}$. Thus,
\beqa
\omega^A(1/2,0,0) &=& \frac{1}{2}\{1+\frac{C_2(1/2)-C_2(0)}{C_2(1/2)}\}=1, \hspace{0.7cm}
\omega^B(1/2,0,0) = \frac{1}{2}\{1+\frac{C_2(0)-C_2(1/2)}{C_2(1/2)}\}=0, \\
\omega^A(1/2,0,1) &=& \frac{1}{2}\{1+\frac{C_2(1/2)-C_2(1)}{C_2(1/2)}\}=-1/3, \hspace{0.7cm}
\omega^B(1/2,0,1)=\frac{1}{2}\{1+\frac{C_2(0)-C_2(1/2)}{C_2(1/2)}\}=0.
\eeqa
We see that the map $T(0,1)$ is useless for cloning.

The case B is straightforward to treat. Suppose now that in the decomposition of
$T$ into irreducible summands, a map $T(1/2,1/2):
\mathscr{B}(\mathscr{H}_{1/2} \otimes \mathscr{H}_{1/2}) \to
\mathscr{B}(\mathscr{H}_{1/2})$ appears. This map decomposes as
$T(1/2)=r(1/2,1/2) T(1/2,1/2)+r(1/2,3/2)T(1/2,3/2)$. The maps
$T(1/2,1/2)$ and $T(1/2,3/2)$ are exactly those encountered in $1
\to 1+1$ cloning. Thus we see immediately from the results of
Section \ref{section:simplest} that: \beqa
\omega^A(1/2,1/2) & = & 1-\frac{4}{3} \lambda(1/2,1/2)^2, \\
\omega^B(1/2,1/2) & = & 1-\frac{4}{3} \mu(1/2,1/2)^2=1-\frac{4}{3}(\frac{\sqrt{3}}{2}
\sqrt{1-\lambda(1/2,1/2)^2}-\frac{1}{2} \lambda(1/2,1/2))^2,
\eeqa
 where $0 \geq |\lambda(1/2,1/2)| \geq 1$, and that the map $T(1/2,3/2)$ is useless for cloning.

We now turn to case C. The convex decomposition of the cloning map
$T$ now contains terms $T(s): \mathscr{B}(\mathscr{H}_{1/2}
\otimes \mathscr{H}_s) \to \mathscr{B}(\mathscr{H}_{1/2})$. Each
of these maps $T(s)$ decomposes as  $T(s)=r(s,s-1) T(s,s-1)+
r(s,s) T(s,s)+ r(s,s+1) T(s,s+1)$.

Each map $T(s,s-1)$ reads 
\beq 
T(s,s-1): \mathscr{O} \to
V(s,s-1)^* (\mathscr{O} \otimes \gras{1}_{s-1}) V(s,s-1), 
\eeq
where 
\beq\label{eq:b9} 
(\pi^A_{1/2} \otimes \pi^B_s \otimes
\pi^X_{s-1})V(s,s-1)=V(s,s-1) \pi_{1/2}. 
\eeq 
There exists a unitary Clebsch-Gordan matrix $C$ such that 
\bed
(\pi^A_{1/2} \otimes (\pi^{BX}_1 \oplus \ldots)) C V(s,s-1)=C V(s,s-1)
\pi_{1/2}. 
\eed
We deduce that 
\beq
\omega^A(s,s-1)=\frac{1}{2}\{1+\frac{C_2(1/2)-C_2(1)}{C_2(1/2)}\}=-1/3.
\eeq 
There also exists a unitary Clebsch-Gordan matrix $D$ such
that 
\bed (\pi^B_{1} \otimes \pi^{AX}_{1/2}) D V(s,s-1)=D V(s,s-1)
\pi_{1/2}, 
\eed if $s=1$, whereas Eq.(\ref{eq:b9}) imply that 
\bed 
(\pi^B_{1} \otimes(\pi^{AX}_{s-1/2} \oplus \pi^{AX}_{s-3/2})) D V(s,s-1)=D V(s,s-1)
\pi_{1/2}, 
\eed
for $s \geq 1$. We infer that 
\beq
\omega^B(s,s-1)=\frac{2}{3}(s+1). 
\eeq

Let us now consider the maps $T(s,s)$. Each such map reads 
\beq
T(s,s): \mathscr{O} \to V(s,s)^* (\mathscr{O} \otimes
\gras{1}_{s}) V(s,s), \eeq where \beq (\pi^A_{1/2} \otimes \pi^B_s
\otimes \pi^X_{s})V(s,s)=V(s,s) \pi_{1/2}. 
\eeq 
Again, there exists some unitary Clebsch-Gordan matrix, which we denote again $C$, such that 
\bed 
(\pi^A_{1/2} \otimes (\pi^{BX}_{0} \oplus \pi^{BX}_{1} \oplus \ldots)) C V(s,s) =C V(s,s) \pi_{1/2}. 
\eed
From Shur's lemma, $V(s,s)$ decomposes as $V(s,s)= \lambda(s,s,0) V(s,s,0) + \lambda(s,s,1) V(s,s,1)$, where $|\lambda(s,s,0)|^2+|\lambda(s,s,1)|^2=1$, where $V(s,s,0)$ intertwines $\pi_{1/2}$ with the (unique) copy of $\pi_{1/2}$ contained in $\pi^A_{1/2} \otimes \pi^{BX}_{0}$ and where $V(s,s,1)$ intertwines $\pi_{1/2}$ with the (unique) copy of $\pi_{1/2}$ contained in $\pi^A_{1/2} \otimes \pi^{BX}_{1}$. Accordingly, we find that 
\beq
\omega^A(s,s)=\frac{1}{2}\{1+\frac{C_2(1/2)-(|\lambda(s,s,0)|^2
C_2(0)+|\lambda(s,s,1)|^2
C_2(1))}{C_2(1/2)}\}=1-\frac{4}{3}|\lambda(s,s,1)|^2, 
\eeq 
A similar reasoning considering the second set of clones gives
$V(s,s)= \mu(s,s,s-1/2) W(s,s,s-1/2) + \mu(s,s,s+1/2)
W(s,s,s+1/2)$, where $|\mu(s,s,s-1/2)|^2+|\mu(s,s,s+1/2)|^2=1$,
where $W(s,s,s-1/2)$ intertwines $\pi_{1/2}$ with the (unique)
copy of $\pi_{1/2}$ contained in $\pi^B_{s} \otimes
\pi^{AX}_{s-1/2}$ and where $V(s,s,s+1/2)$ intertwines $\pi_{1/2}$
with the (unique) copy of $\pi_{1/2}$ contained in $\pi^B_{s}
\otimes \pi^{AX}_{s+1/2}$. Accordingly, \bed
\omega^B(s,s)=\frac{1}{2}\{1+\frac{C_2(s)-(|\mu(s,s,s-1/2)|^2
C_2(s-1/2)+|\mu(s,s,s+1/2)|^2 C_2(s+1/2))}{C_2(1/2)}\}, \eed The
intertwiners $V$'s and $W$'s are explicitly given by \beqa &&
V(s,s,0)=C^{(1/2 u)(0 0)}_{(1/2 m)}C^{(s v )(s w)}_{(0 0)}
\ket{1/2,u}_A \ket{s,v}_B \ket{s,w}_X
\bra{1/2,m},\\
&& V(s,s,1)=C^{(1/2 u )(1 l)}_{(1/2 m)}C^{(s v)(s w)}_{(1 l)}
\ket{1/2,u}_A \ket{s,v}_B \ket{s,w}_X
 \bra{1/2,m},\\
&& W(s,s,s-1/2)=C^{(s v)(s-1/2 z)}_{(1/2 m)}C^{(1/2 u)(s
w)}_{(s-1/2 z)} \ket{1/2,u}_A \ket{s,v}_B
\ket{s,w}_X  \bra{1/2,m},\\
&& W(s,s,s+1/2)=C^{(s v)(s+1/2 y)}_{(1/2 m)}C^{(1/2 u)(s
w)}_{(s+1/2 y)} \ket{1/2,u}_A \ket{s,v}_B \ket{s,w}_X \bra{1/2,m}.
\eeqa The following relations hold between the $\lambda$'s and the
$\mu$' s: \beqa
\mu(s,s,s-1/2) &= & \frac{1}{2}\{ \lambda(s,s,0) \tr W(s,s-1/2)^*
V(s,s,0)+\lambda(s,s,1) \tr W(s,s,s-1/2)^* V(s,s,1)\} \\
\mu(s,s,s+1/2) &=& \frac{1}{2}\{ \lambda(s,s,0) \tr W(s,s,s+1/2)^*
V(s,s,0) +\lambda(s,s,1) \tr W(s,s,s+1/2)^* V(s,s,1)\}.
\eeqa

From an explicit calculation(using Mathematica), one gets:
\beqa
\tr W(s,s,s-1/2)^* V(s,s,0) &= & -\tr W(s,s,s+1/2)^*
V(s,s,1)=2 \sqrt{\frac{s}{2(s+1/2)}},\\
\tr W(s,s,s-1/2)^* V(s,s,1) &= & \tr W(s,s,s+1/2)^*
V(s,s,0)=\sqrt{\frac{2(s+1)}{s+1/2}}. \eeqa
From which we find
\bed
\omega^B(s,s)=\frac{2}{3}(1-|\lambda(s,s,0)|^2)+\frac{4}{3}\lambda(s,s,0)\lambda(s,s,1)\sqrt{s(s+1)}.
\eed

We now turn to the third and last piece: the maps
$T(s,s+1):\mathscr{B}(\mathscr{H}_{1/2} \otimes \mathscr{H}_{s})
\to \mathscr{B}(\mathscr{H}_{1/2})$. With a reasoning similar to
the analysis of the maps $T(s,s-1)$ and $T(s,s)$, one finds that
\bed
\omega^A(s,s+1)=\frac{1}{2}\{1+\frac{C_2(1/2)-C_2(1)}{C_2(1/2)}\}=-1/3.
\eed and \bed
\omega^B(s,s+1)=\frac{1}{2}\{1+\frac{C_2(s)-C_2(s+1/2)}{C_2(1/2)}\}=-\frac{2}{3}s.
\eed So, we see that the maps $T(s,s+1)$ are useless for cloning.

\textbf{Extremisation}. Let us first consider the case where $n$ is even. Optimal cloning maps are of the form 
\bed 
T=r(0,0)T(0,0)+\sum_{s=1}^{n/2}(r(s,s-1)T(s,s-1)+r(s,s)T(s,s)). 
\eed 
Note that $\omega^A(s,s-1)=-1/3 \; \forall s \geq 1$ and that $\omega^B(s,s-1)$ strictly increases with $s$. Thus, for the sake of optimality, we can choose $r(s,s-1)=0 \; \forall s < n/2$. Also, one can see that optimal maps can be found for $\lambda(s,s,0), \lambda(s,s,1) \geq 0$ and for $r(s,s)=0 \; \forall s<n/2$. Introducing lighter notations; $r(0,0) \equiv a$, $r(n/2,n/2-1) \equiv b$, $r(n/2,n/2) \equiv c$, $\lambda(n/2,n/2,0) \equiv x$, and $\lambda(n/2,n/2,1) \equiv y$, and extremising,  we get Eqs.(\ref{eq:onetoneven}).

A similar argument holds when $n$ is odd ($n \geq 3$) and also leads to Eqs.(\ref{eq:onetoneven}).

\end{document}